\documentclass[useAMS,usenatbib]{mnras}
\input psfig.sty

\def \aap  {A\&A}

\def \aj  {AJ}
\def \apj  {ApJ}
\def \apjs  {ApJS}
\def \apjl  {ApJL}

\def \jcap {Journal of Cosmology and Astroparticle Physics}
\def \prd {Phy.Rev.D}
\def \pasp {PASP}
\def \mnras {MNRAS}
\def \nat {Nature}
\def \na {New Astronomy}
\def \etal {et~al.~}
\def \chisq  {\ifmmode  \chi^2   \else  $\chi^2$  \fi}  
\def \spose#1{\hbox  to 0pt{#1\hss}}  
\def \lta{\mathrel{\spose{\lower 3pt\hbox{$\sim$}}\raise  2.0pt\hbox{$<$}}}
\def \gta{\mathrel{\spose{\lower  3pt\hbox{$\sim$}}\raise 2.0pt\hbox{$>$}}}

\def \kms {\ifmmode  \,\rm km\,s^{-1} \else $\,\rm km\,s^{-1}  $ \fi }
\def \kpc {\ifmmode  {\rm~kpc}  \else ${\rm~kpc}$\fi}  
\def \pc {\ifmmode  {\rm~pc}  \else ${\rm~pc}$ \fi  }  
\def \Gyr {\ifmmode  {\rm~Gyr}  \else ${\rm~Gyr}$\fi}
\def \Msun {\ifmmode {\rm M}_{\odot} \else M$_{\odot}$ \fi} 
\def \Lsun {\ifmmode {\rm L}_{\odot} \else L$_{\odot}$ \fi} 
\def \Rsun {\ifmmode {\rm R}_{\odot} \else R$_{\odot}$ \fi} 
\def \Msunpyr {\ifmmode {\rm M}_{\odot}{\rm~yr}^{-1} \else M$_{\odot}{\rm~yr}^{-1}$ \fi} 

\def \LCDM {\ifmmode \Lambda{\rm CDM} \else $\Lambda{\rm CDM}$ \fi}
\def \sig8 {\ifmmode \sigma_8 \else $\sigma_8$ \fi} 
\def \OmegaM {\ifmmode \Omega_{\rm m} \else $\Omega_{\rm m}$ \fi} 
\def \Omegab {\ifmmode \Omega_{\rm b} \else $\Omega_{\rm b}$ \fi} 
\def \OmegaL {\ifmmode \Omega_{\rm \Lambda} \else $\Omega_{\rm \Lambda}$\fi} 
\def \Deltavir {\ifmmode \Delta_{\rm vir} \else $\Delta_{\rm vir}$ \fi}
\def \rhocrit {\ifmmode \rho_{\rm crit} \else $\rho_{\rm crit}$ \fi}
\def \rhou {\ifmmode \rho_{\rm u} \else $\rho_{\rm u}$ \fi}

\newcommand{\hMpc}{{\ifmmode{h^{-1}{\rm Mpc}}\else{$h^{-1}$Mpc}\fi}}
\newcommand{\hkpc}{{\ifmmode{h^{-1}{\rm kpc}}\else{$h^{-1}$kpc}\fi}}
\newcommand{\Rvir}{{\ifmmode{R_{\rm vir}}\else{$R_{\rm vir}$}\fi}}
\def \Mvir {\ifmmode M_{\rm  vir} \else $M_{\rm  vir}$ \fi}  
\newcommand{\Mstar}{{\ifmmode{M_{\star}}\else{$M_{\star}$}\fi}}
\newcommand{\Vrot}{{\ifmmode{V_{\rm rot}}\else{$V_{\rm rot}$}\fi}}

\newcommand{\ltsima}{$\; \buildrel < \over \sim \;$}
\newcommand{\gtsima}{$\; \buildrel > \over \sim \;$}
\newcommand{\lsim}{\lower.5ex\hbox{\ltsima}}
\newcommand{\gsim}{\lower.5ex\hbox{\gtsima}}

\title[Dark matter properties in the NIHAO simulations]{NIHAO project
  II: Halo shape, phase-space density and velocity distribution of
  dark matter in galaxy formation simulations} 

\author[Butsky \etal]{Iryna Butsky$^{1,2,3}$\thanks{E-mail: ibutsky@uw.edu},
Andrea V. Macci{\`o}$^{4,1}$\thanks{E-mail: maccio@nyu.edu},   Aaron A. Dutton$^{4,1}$, Liang Wang$^{5,1,6}$, 
\newauthor Aura Obreja$^4$, Greg S. Stinson$^1$, Camilla  Penzo$^{7,1}$, Xi Kang$^5$, Ben W. Keller$^8$,
\newauthor James Wadsley$^8$\\
$^1$ Max-Planck-Institut f\"ur Astronomie, K\"onigstuhl 17, 69117 Heidelberg, Germany\\
$^2$ University of Washington, Seattle, Washington, USA \\
$^3$ California Institute of Technology, Pasadena, California, USA \\ 
$^4$ New York University Abu Dhabi, PO Box 129188, Abu Dhabi, UAE \\
$^5$ Purple Mountain Observatory, the Partner Group of MPI f\"ur Astronomie, 2 West Beijing
Road, Nanjing 210008, China\\
$^6$ Chinese Academy of Science Graduate School\\
$^7$ Laboratoire\,Univers\,et\,Th\'eories,\,UMR\,8102\,CNRS,\,Observatoire\,de\,Paris,\,Universit\'e\,Paris Diderot,\,5 Place\,Jules\,Janssen,\,92190\,Meudon,\,France \\
$^8$ Department of Physics and Astronomy, McMaster University, Hamilton, Ontario L8S 4M1, Canada}

\begin{document} 
              
\date{\today}
              
\pagerange{\pageref{firstpage}--\pageref{lastpage}}\pubyear{2016} 
 
\maketitle 

\label{firstpage}
              
\begin{abstract}
We use the NIHAO (Numerical Investigation of Hundred Astrophysical
Objects) cosmological simulations to study the effects of galaxy
formation  on key  properties of dark matter (DM) haloes. NIHAO
consists of $\approx 90$ high-resolution SPH simulations that include
(metal-line) cooling, star formation, and feedback from massive stars
and SuperNovae, and cover a wide stellar and halo mass range: $10^6
\lta M_* / \Msun \lta 10^{11}$ ( $10^{9.5} \lta M_{\rm halo} / \Msun
\lta 10^{12.5}$).
When compared to DM-only simulations, the NIHAO haloes have similar
shapes at the virial radius, \Rvir, but are substantially rounder
inside $\approx 0.1\Rvir$.  In NIHAO simulations $c/a$ increases with
halo mass and integrated star formation efficiency, reaching $\sim
0.8$ at the Milky Way mass (compared to 0.5 in DM-only), providing a
plausible solution to the long-standing conflict between observations
and DM-only simulations.
The radial profile of the phase-space $Q$ parameter ($\rho/\sigma^3$)
is best fit with a single power law in DM-only simulations, but
shows a flattening within $\approx 0.1\Rvir$ for NIHAO for total
masses $M>10^{11} \Msun$.
Finally, the global velocity distribution of DM is similar in both
DM-only and NIHAO simulations, but in the solar neighborhood, NIHAO
galaxies deviate substantially from Maxwellian. The distribution is
more symmetric, roughly Gaussian, with a peak that shifts to higher
velocities for Milky Way mass haloes. We provide the distribution
parameters which can be used for predictions for direct DM detection
experiments.
Our results underline the ability of the galaxy
formation processes to modify the properties of dark matter haloes.
\end{abstract}


\setcounter{footnote}{1}

\section{Introduction}
\label{sec:intro}

N-body numerical cosmological simulations have proven to be powerful
tools for modeling the properties of the 3-dimensional dark matter
distribution in galactic haloes \citep{JingSuto2002, Allgood2006,
  Bett2007, Hayashi2007,  Maccio2007, Neto2007}.  However, these
dark-matter-only cosmological  simulations only take into account the
effects of gravity.  Globally, the baryons  only make up $\Omega_{\rm
  b}/\Omega_{\rm m} \approx 15\%$ of the mass  budget
\citep{Planck2014}, and thus on large scales their influence can be
reasonably ignored.   However, baryons can dissipate energy, leading
to large mass fractions near the centers of galaxies. Additional
processes such as the violent explosion of stars as supernovae can
have non-negligible impact on the dark matter distribution.

Using the $\Lambda$CDM cosmology,   dark-matter-only galaxy
simulations can make precise predictions of the shape and the internal
matter and velocity distribution of dark matter haloes
\citep{Maccio2008}. On average, such dissipationless cosmological
simulations predict the ratio of the minor principal axis to the major
to be $\langle c/a \rangle\sim0.6$.  Observations, however, indicate
that haloes are more spherical. \citet{Ibata2001} used the Sagittarius
Stream (and a \cite{DehnenBinney1998} halo mass distribution)
to  put a lower bound on the shape of the inner Milky Way halo
of $c/a \geq 0.8$, where the halo was defined to be between $20 $ kpc
$ < r < 60$ kpc.  Other recent work found similarly spherical shapes
for the inner halo of the Milky Way \citep{Helmi2004, Martinez2004}.

In a recent, detailed study, \citet{Law2010} presented a new N-body
model (with a logarithm dark matter halo)\footnote{An NFW profile will
  give similar results \citep{Law2010}} for the tidal disruption of
the Sagittarius galaxy which led them to conclude that our Galaxy has
an oblate potential with a minor to major axis ratio of
$(c/a)_{\phi}=0.72$. This latter value can be seen as a upper limit on
the shape of the matter distribution, which again suggests a possible
tension between results of pure CDM simulations and local measurements
of halo shape.

One possible solution to this problem could be related to the
inclusion of a dissipative component such as baryons in cosmological
simulations.  \citet{Katz1991} and \citet{Dubinski1994} were among the
first to show that haloes in dissipative simulations were
systematically more spherical than corresponding haloes in
dissipationless simulations.  \citet{Kazantzidis2004} analyzed the effect of
baryons on the shape of dark matter haloes using high resolution
hydrodynamical simulations of clusters of galaxies and, again, found a
reduced triaxiality in dissipational simulations.  That result was
confirmed in several further studies \citep{Springel2004, Debattista2008,
  Pedrosa2009, Tissera2010}.  While there is an agreement that baryons tend to
make dark matter haloes rounder, a quantitative prediction of this
effect strongly depends on the prescription used to model baryons in
cosmological simulations, which is not a trivial task.

A large fraction of simulated galaxies used for studying the impact of
baryons on the DM distribution were either plagued by the so called
over-cooling problem \citep[which produced galaxies that were
  unrealistically dominated by baryons in their central
  region][]{Kazantzidis2004} or were missing important aspects of the galaxy
formation process such as star formation and feedback \citep{Abadi2010}.

A noticeable exception has been the recent work by \citet{Bryan2013},
which employed the OWLs simulations \citep{Schaye2010}, a large scale
simulation that produced realistic large galaxies and cosmic structure
in a full cosmological framework.  These simulations found that
baryons substantially changed $c/a$ of massive galaxies.  On the other
hand they were able to resolve galaxies on the scale of the Milky Way
with only few thousand particles.

Shape is not the only parameter of dark matter haloes that baryons
might modify. A lot of recent attention has focused on the
modification (expansion and core creation) of density profiles \citep{Gnedin2004, Abadi2010,
  Governato2010, Maccio2012, Pontzen2012, DiCintio2014b, Dutton2015,
  Onorbe2015}, which helps reduce the tension on small-scales between
observations \citep[e.g.][and references therein]{Oh2015}
and DM-only predictions \citep{Tollet2016, Dutton2016}.
Another interesting quantity that is a slight variant on the density
profile is the so called coarse-grained phase-space density
$Q=\rho/\sigma^3$ that takes into account particle velocities as well
as matter density.  DM-only simulations show that radial profiles of
$Q$ are well fit with a simple power law $\rho/\sigma^3 \propto r^{\alpha}$
\citep{Taylor2001}.

Analytical work found a characteristic value for the power law slope
of the Q profile, $\alpha = 1.944$, for isotropic structures
\citep{Austin2005}. This value also reproduces realistic radial velocity
dispersions through cosmological simulations (\citet{Dehnen2005}, but see
also \citet{Schmidt2008} for a more critical interpretation).  It is
particularly interesting to check if baryons break this simple power
law behavior of the dark matter coarse-grained phase-space density in
simulated galaxies.

The dark matter {\it velocity} distribution is another very important
parameter, since it has profound implications for the predicted dark
matter - nucleon scattering rates in direct detection experiments
\citep{Kuhlen2010}.  Most models assume a Maxwell-Boltzmann (MB) velocity
distribution function.  Therefore, a departure from the MB
distribution might change the predicted rate of events for dark matter
models in which the scattering is sensitive to the high velocity tail
of the distribution, and for experiments that require high energy
recoil events \citep[e.g.][]{Vergados2008}.

Recent high resolution DM-only simulations have reported substantial
departures from a Maxwellian shape \citep{Vogelsberger2009, Kuhlen2010}, with
a deficit near the peak and excess particles at high speeds.   In
contrast, an analysis of the ERIS hydrodynamical simulation of a
single Milky Way like galaxy, showed a different behavior.  Its
velocity function maintains the Maxwellian distribution and  shows a
greater deficit than MB at high velocities \citep{Pillepich2014}.

In this paper we revisit the issues of the effect of baryons on
dark matter properties using the galaxies from the NIHAO project
\citep{Wang2015}. 
The NIHAO project is a large suite of high
resolution simulated galaxies. The sample includes more than 90 high
resolution zoom regions with  halo masses ranging from $\sim10^{10}$
to $\sim10^{12}\Msun$.  Each high resolution galaxy is resolved with
at least $600,000$ elements (DM+GAS) and up to several millions [see
\citet{Wang2015} for details].  It is the largest sample of high
resolution galaxies to date.  The simulations show remarkable
agreement with the stellar mass--halo mass relationship across five
orders of magnitude of stellar mass.  Thus a fully sampled volume
would follow the  observed galaxy stellar mass function.  The
simulations thus offer a unique tool to study the modifications that
baryons induce to dark matter since it simultaneously combines high
spatial and mass resolution with a statistical sample of galaxies
across three orders of magnitude in halo mass.

We have decided to focus this paper on three
main properties: the halo shape, since it is a widely studied halo
property and it is possible to directly measure it, especially
in our own galaxy. The (pseudo) phase-space density, since
it has been claimed to be an universal property of DM haloes
\citep{Taylor2001} and it is somehow complementary to the
DM density profiles which have been discussed in the NIHAO IV paper,
and the DM velocity distribution that has important consequences
on the predicted rate of events in direct detection experiments.
Another interesting property is the angular momentum distribution
of the DM, but this will be the topic of a different NIHAO paper
more closely linked to the process of disc formation (Obreja
\etal in prep.)

This paper is organized as follows: in Section 2 we discuss the
simulations including the feedback  prescription used in creating the
NIHAO galaxies; In Section 3 we discuss in detail the effects of
baryons on the dark matter halo shape, pseudo phase space density and
the velocity distribution. Section 4 offers a summary of our results,
as well as their impact.

\section{Simulations} \label{sec:sims}

\setcounter{table}{0}
\begin{table*}
\begin{minipage}{180mm}
\begin{center}
  \caption{Properties of the four selected ``test'' galaxies. $\Mstar$
    is the total stellar mass within $\Rvir$. Note that in the NIHAO
    project the ``name'' of a galaxy indicates the mass of the halo in
    the DM-only simulation.}
\label{tab:gal}
\begin{tabular}{lccccccccc}
\hline
Simulation & $\Mstar$ &$\Rvir$& Softening & $m_{\rm DM}$ & $m_{\rm gas}$ & $N_{\rm vir}^{\rm DM}$ & $N_{\rm vir}^{\rm *}$ & Mass bin & Mass range\\ 
name  &[M$_\odot$]& [kpc]& [pc] & [M$_\odot$] &  [M$_\odot$] &  & & name  &[M$_\odot$]\\

\hline
g1.92e12 &  1.59e+11 & 281 & 931 & 1.74e+06 & 3.16e+05 & 1,200,667 & 2,467,821   & M3 & 7.5e+11$<M_{\rm vir}<$3.5e+12\\ 

g5.02e11 &  1.46e+10 & 179 & 465 & 2.17e+05 & 3.95e+04 & 2,408,247 & 1,821,403   & M2 & 2.5e+11$<M_{\rm vir}<$7.5e+11\\ 

g1.08e11 &  8.47e+08 & 105 & 465 & 2.17e+05 & 3.95e+04 & $\phantom{,1}$520,108 & $\phantom{,1}$108,025   & M1 & 7.5e+10$<M_{\rm vir}<$2.5e+11\\ 

g4.99e10 &  1.24e+08 & $\phantom 1$78 & 207 & 1.90e+04 & 3.48e+03 & $\phantom{,1}$732,808 & $\phantom{,11}$52,511     & M0 & 4.0e+09$<M_{\rm vir}<$7.5e+10\\ 

\hline  
\end{tabular}
\end{center}
\end{minipage}
\end{table*}

We study simulations from the NIHAO project \citep{Wang2015}, that use
baryonic physics from an updated version of the MaGICC simulations
\citep{magic}.  The initial conditions are created using the latest
compilation of cosmological parameters from the Planck satellite
\citep{Planck2014}; namely $\OmegaL=0.6825$, \OmegaM=0.3175,
$H_0=67.1$, $\sigma_8=0.8344$, $n=0.9624$ and $\Omegab=0.0490$.  The
haloes to be re-simulated with baryons and higher resolution were
extracted from three different pure N-body simulations with box sizes
of 60, 20 and 15 \hMpc, more information on the DM-only simulations
can be found in \citet{Dutton2014}.

The NIHAO project was designed to study galaxy formation over a wide
mass range, from dwarf galaxies to massive spirals like the Milky-Way.
The simulations maintain the same {\it numerical resolution} across
the entire mass range. This means that all haloes are resolved
 with at least 600,000 elements (DM+stars+GAS) within the virial
 radius, up to several millions. This fixes numerical resolution
 has been achieved by varying both the size of the initial
 box from which the haloes to be zoomed have been chosen, and the number
 of zoom levels of each simulation (see figure 2 in \citet{Wang2015}).

The zoomed initial conditions were created using a modified version of
{\sc grafic2} \citep{Bertschinger2001, penzo2014}.  All haloes/galaxies
presented in this paper are the most massive object in their
respective high resolution volume, in other words we will only present
results, at all masses, for central haloes.  The starting redshift is
$z_{\rm start}=100$, and each halo is initially simulated at high
resolution with DM-only using {\sc pkdgrav} \citep{Stadel2001}.  More
details on the sample selection can be found in \citet{Wang2015}.

We refer to  simulations with baryons as the {\it Hydro} simulations or
NIHAO, while we will use the term {\it N-body} or DM-only for the collisionless simulations.

The hydro simulations are evolved using an improved version of the SPH
code {\sc gasoline} \citep{gasoline}.  The code includes a subgrid
model for turbulent mixing of metals and energy \citep{Wadsley2008},
heating and cooling include photoelectric heating of dust grains,
ultraviolet (UV) heating and ionization and cooling due to hydrogen,
helium and metals \citep{Shen2010}.

For the NIHAO simulations we have used a revised treatment of
hydrodynamics described in \citet{Keller2014} that we refer to as {\sc
  esf-Gasoline2}.  Most important is the \citet{Ritchie2001} force
expression that improves mixing and shortens the destruction time for
cold blobs (see \citet{Agertz2007}).  {\sc esf-Gasoline2} also includes
the time-step limiter suggested by \citet{Saitoh2009}, which is important
in the presence of strong shocks and temperature jumps.  We also
increased the number of neighbor particles used in the calculation of
the smoothed hydrodynamic properties from 32 to 50.

 \begin{figure}
\centerline{\psfig{figure=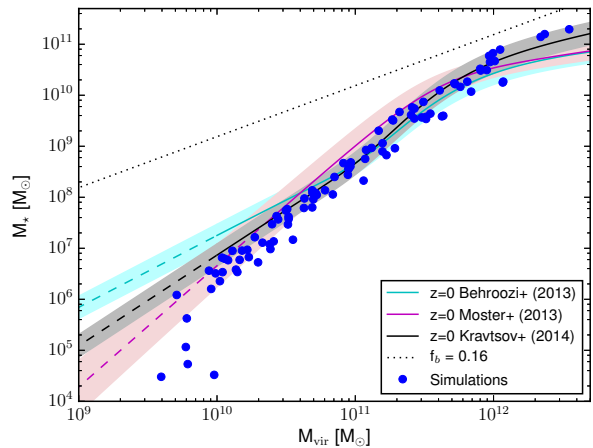,width= 0.49\textwidth}}
  \caption{\scriptsize Stellar mass - halo mass relation for the NIHAO
    galaxies used in this work. All simulations have more than 600,000
    particles in their virial radius (see Wang \etal 2015 for more
    details). The solid lines represent the most commonly used
    abundance matching results (see text).}
\label{fig:msmv}
\end{figure}

\subsection{Star formation and feedback}

The star formation and feedback modeling follows what was used in the
MaGICC simulations \citep{magic}.  Gas can form stars  when it
satisfies a temperature and a density threshold: $T< 15000$ K and
$n_{\rm th} > 10.3$ cm$^{-3}$.  Stars can feed energy back into the
ISM via blast-wave supernova (SN) feedback \citep{Stinson2006} and via
ionizing radiation from massive stars before they turn in SN.  Metals
are produced by type II SN, type Ia SN.  These, along with stellar
winds  from asymptotic giant branch stars also return mass to the ISM.
The metals affect  the cooling function \citep{Shen2010}  and diffuse
between gas particles \citep{Wadsley2008}.  The fraction of stellar
mass that results in SN and winds is determined using  the
\citet{Chabrier2003} stellar Initial Mass Function (IMF).  

There are two small changes from the MaGICC simulations.   The change
in number of neighbors and the new combination of softening length and
particle mass means the threshold for star formation increased from
9.3 to 10.3 cm$^{-3}$,  The increased hydrodynamic mixing necessitated
a small increase of pre-SN  feedback efficiency, $\epsilon_{\rm ESF}$,
from 0.1 to 0.13.   This energy is ejected as thermal energy into the
surrounding gas, which does not have its cooling disabled.  Most of
this energy is instantaneously radiated away, and the effective
coupling is of the order of 1\%.  

 \begin{figure}
\centerline{\psfig{figure=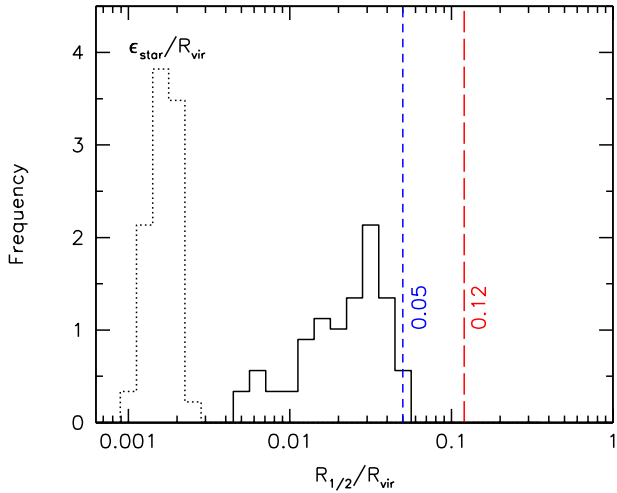,width= 0.45\textwidth}}
   \caption{\scriptsize Histogram of the ratio between stellar half
     mass radius, $R_{1/2}$, and the viral radius, $\Rvir$.  For
     reference, the three radii at which we will  compute the halo
     shape are: $\Rvir$, 12\% $\Rvir$ (red line)  and 5\% $\Rvir$
     (blue line). The leftmost histogram shows the distribution of the
   stellar particle softening through NIHAO.}
\label{fig:msrad}
\end{figure}

\subsection{The galaxy sample}

For this project we use a total of 93 galaxies from the NIHAO sample,
all galaxies are ``centrals'' in their respective haloes and are
resolved with have more than 600,000 particles (dm+gas+stars) within
the virial radius.\footnote{In this paper we have used the latest
  compilation of NIHAO galaxies, including several new ``central''
  galaxies that were completed after the publication of the first
  NIHAO paper.}  Since our aim is to study the impact of galaxy
formation on the dark matter distribution it is very important to use
realistic simulated galaxies, i.e. galaxies able to reproduce the
observed scaling relations.  One of the key success of the NIHAO
galaxies is to be able to reproduce the observed stellar mass-halo
mass relation across the whole mass spectrum, which extends for more
than 5 orders of magnitude in stellar mass, as shown in
Fig.~\ref{fig:msmv}.  As detailed in \citet{Wang2015} the NIHAO
galaxies are also able to reproduce the the stellar mass - halo mass
relation also at higher redshift, and have realistic star formation
rates for their stellar masses.  NIHAO galaxies are also consistent
with the observed gas content of galaxy discs and  haloes
\citep{Stinson2015,Wang2016,Gutcke2016}.  Thanks to the unprecedented
combination of high resolution and large statistical sample, the NIHAO
suite offers a unique tool to study the distribution response of
several DM properties to galaxy formation.

Since among our goals there is the study of the shape of the DM
distribution at different radii, it is interesting to check how these
radii compare to  the size of the central stellar region.
Fig.~\ref{fig:msrad} shows a histogram of the ratio between the
stellar half-mass radius, $R_{1/2}$, and the viral radius, $\Rvir$.
The three key radii at which we will compute the halo shape are:
\Rvir, 12\% $\Rvir$  and 5\% \Rvir.  As the plot shows all these radii
are larger than the size of the stellar body and are well above the
softening of the simulations which is shown by the leftmost histogram
in the figure (see also Table~\ref{tab:gal} and Wang \etal
2015). Finally we have decided to use as minimum radius for our
profiles 2\% of \Rvir, which again is well within the softening of all
simulations.  A more detailed discussion of the ``convergence radius''
of our simulations can be found in \cite{Tollet2016}.

\begin{figure*}
\centerline{\psfig{figure=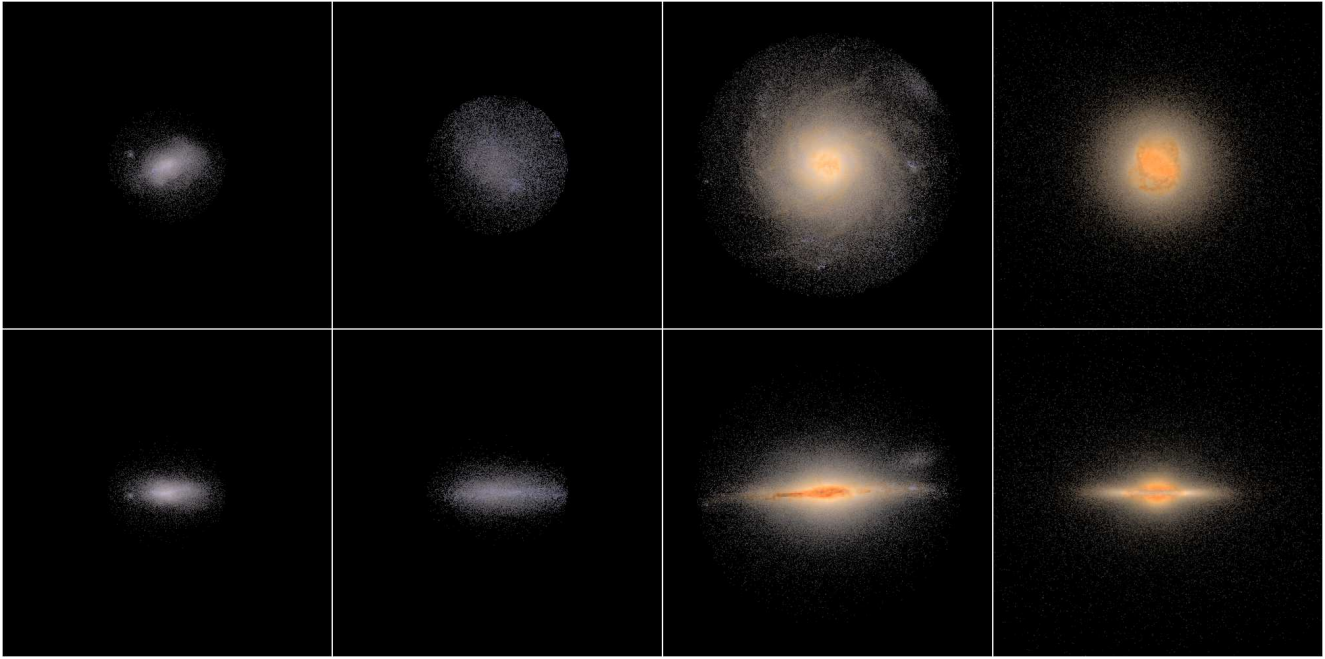,width= 0.98\textwidth}}
\caption{\scriptsize Face-on (upper panels) and edge-on (lower panels)
  views of our four selected ``test'' galaxies: g4.99e10, g1.08e11,
  g5.02e11, g1.92e12 from left to right.  Galaxies have been processed
  through the Monte Carlo radiative transfer code {\sc sunrise}.
  Images are 50 kpc on a side.}
\label{fig:4test}
\end{figure*}

In order to organize our results we have decided to show detailed
results (as for example the triaxiality parameter as a function of
radius) for only four ``test'' galaxies. These galaxies have masses
equally spaced by half a dex ranging from $5\times10^{10}$ to
$10^{12}$ and are listed in Table~\ref{tab:gal}. A visual impression
of our galaxies is shown in figure \ref{fig:4test}, and was created
using the Monte Carlo radiative transfer code {\sc sunrise}
\citep{Jonsson2006}. The image brightness and contrast are scaled
using arcsinh as described in \citet{Lupton2004}.  Around the mass
each of those galaxies we have then constructed a (total) mass bin of
size of 0.5 dex (i.e. the mass bin around g1.08e11 goes from
$7.5\times 10^{10}$ to $2.5\times 10^{11} \Msun$). No other selection
criterion is used besides the mass.  In the following, for all
inspected DM properties (shape, phase-space and velocity distribution)
we will present results for our single test galaxies and then average
results for the 4 mass bins: M0-M3, listed in increasing mass order.

\section{Results}

To determine the properties of the dark matter halo we considered all
dark matter particles within $\Rvir$ (the radius that encompasses a
density equal to 200 times the critical density of the Universe) found
using the Amiga Halo Finder (AHF) \citep{ahf}.  The study makes a
direct comparison between the haloes that form in the DM-only
simulations (black symbols in each Fig.)  and those that form in the
hydro simulations (red symbols).

The comparison is made between the axis ratios ($b/a$, $c/a$ and $T$
the triaxiality), the pseudo phase space density ($Q$), and the
velocity distribution.  We investigate the velocity dispersion both
globally and in the Solar neighborhood.

 \begin{figure*}
\centerline{\psfig{figure=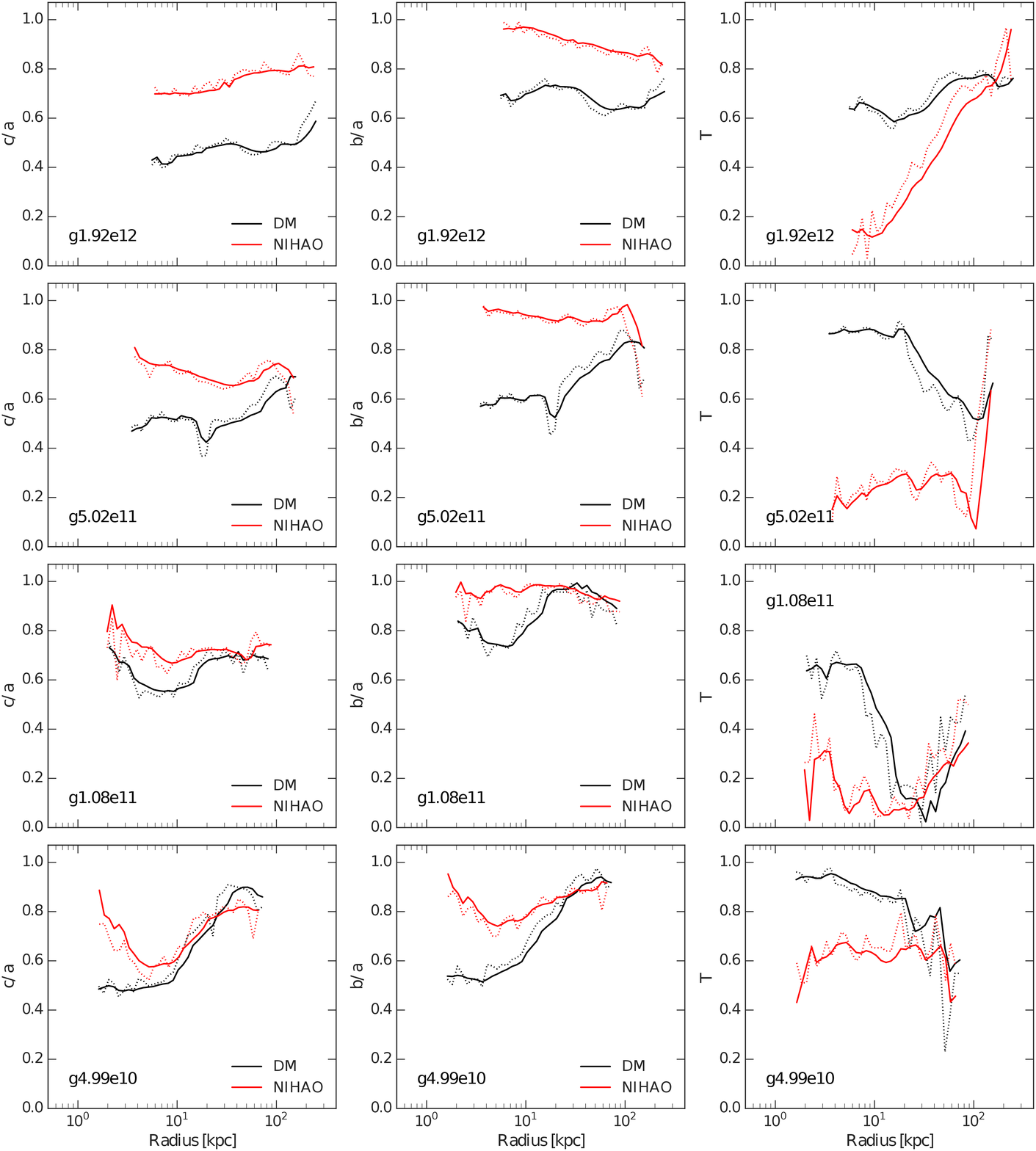,width=1.0\textwidth}}
  \caption{\scriptsize Ratio of minor-to-major axes ($c/a$, left),
    middle-to-major axes ($b/a$, center) and triaxiality ($T$, right)
    as a function of radius. DM-only simulations are depicted in black
    while hydro (NIHAO) simulations are depicted in red.  Both the
    integral (solid line) and differential (dotted line) values are
    shown.  }
\label{fig:shape1}
\end{figure*}

 \begin{figure*}
\centerline{\psfig{figure=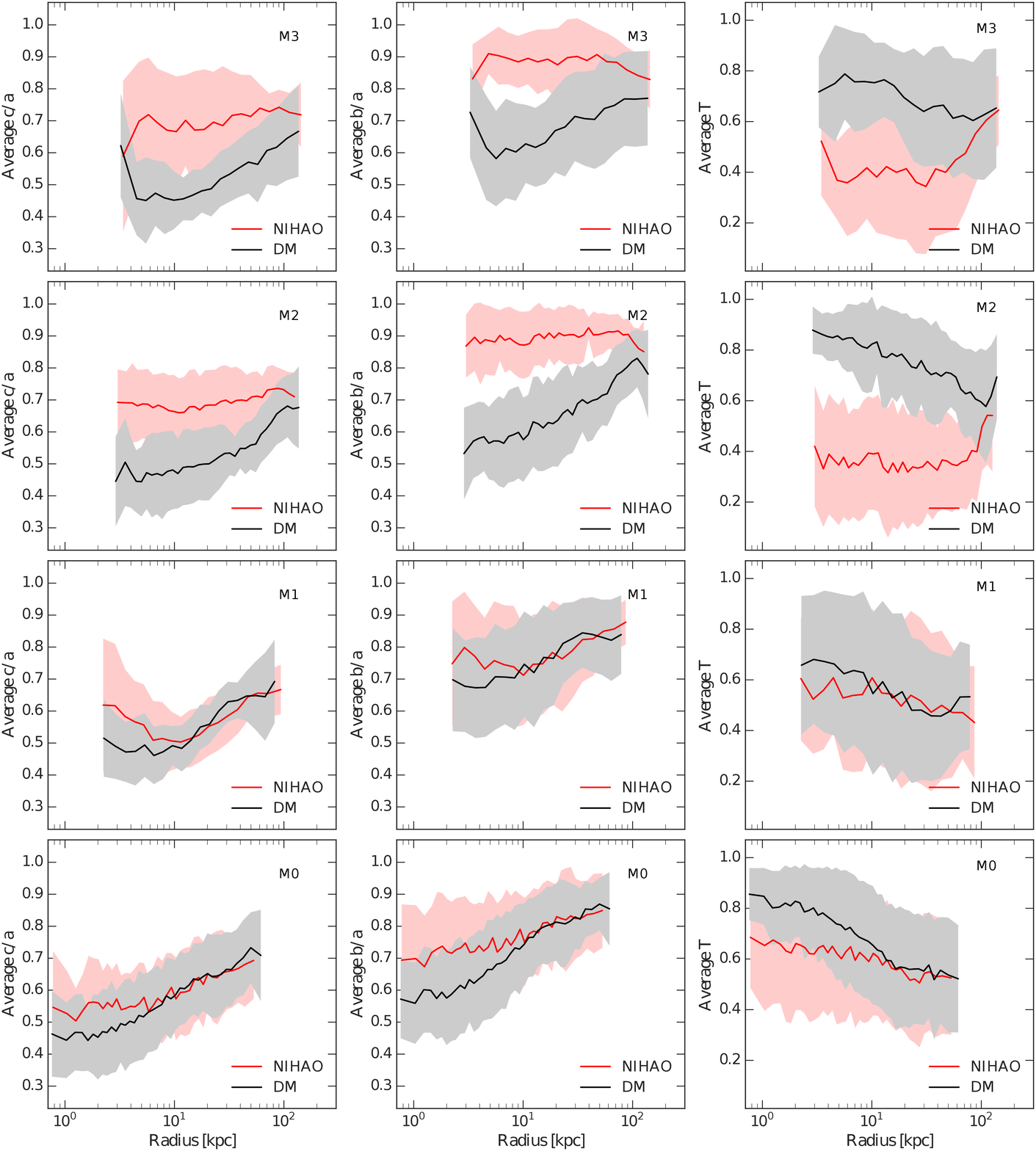,width=1.0\textwidth}}
  \caption{\scriptsize Same as Fig.~\ref{fig:shape1} for all galaxies
    in the four  mass bins M3-M0 (M0 $<$ M1 $<$ M2 $<$ M3 $\sim
    10^{12}\Msun$).  The shaded area represents the 1$\sigma$ scatter
    from galaxy to galaxy in the respective mass bin. Only the
    integral shape (solid line) is shown. }
\label{fig:shape1_all}
\end{figure*}

\subsection{Halo shape}

For each simulation (DM-only and hydro), we define the principal axes
of the shape ellipsoid such that $a > b > c$ and define the ratio
parameters, as $s = c/a, q = b/a, p = c/b$.  To calculate the shape of
the dark matter within a given radius, we  begin with the assumption
of a spherical ellipsoid, such that $a = b = c$.  We then compute the
inertia tensor $I_{ij}$, defined as: 
\begin{center}
$I_{ij} = \Sigma _{\alpha} m_{\alpha}x_i^{\alpha}x_j^{\alpha}/r_{\alpha}^2$,
\end{center}
where $m$ is the particle mass, $\alpha$ is the particle index and
$i,j$ refer to the coordinates.  The radius is defined to be
$r_{\alpha}^2 = x_{\alpha}^2 + y_{\alpha}^2/q^2 + z_{\alpha}^2/s^2$
\citep{Kazantzidis2004}.

The eigenvalues of this matrix produce new values for $s, q, p$. We
iterate,  using the eigenvalues of the previous matrix as the new
assumptions of $s$ and $q$, until the fractional difference of $s, q, p$
converges to a tolerance value of $10^{-3}$ \citep{Maccio2008}.

Finally we define the triaxiality parameter as $T=[1 - (b/a)^2] / [1
  -(c/a)^2]$.  A prolate halo has $T=1$, and oblate halo has $T=0$,
while a triaxial halo has $T\sim 0.5$.

In Fig.~\ref{fig:shape1}, we present the relation between the minor to
major axis ratio ($c/a$, left) the middle to major ($b/a$, center) and
the triaxiality parameter (right)  as a function of radius.  The solid
line is for the integral measurement of the shape (i.e. $c/a(<r)$)
while the dotted line is for the differential one.  In each panel the
last point marks the virial radius of the halo.  As discussed
previously for clarity and brevity we present results only for our 4
test galaxies (see Table~\ref{tab:gal}) out of 93 in our sample,

The trend is similar across all masses: baryonic physics processes
such as cooling  and stellar feedback modify the shape of the DM
distribution.  The effect of baryons physics is strongest at the
center of the halo.  In general, the hydro simulations have a less
varying shape as a function of radius w.r.t Nboby sims and the hydro
shape is always rounder than the dark matter.  At small radii, the
dark matter becomes more triaxial.  Near \Rvir, the shapes of the
hydro and DM-only simulations converge to their most spherical shape
in all the galaxies except  the most massive one, g1.92e12.

\begin{figure*}
\centerline{\psfig{figure=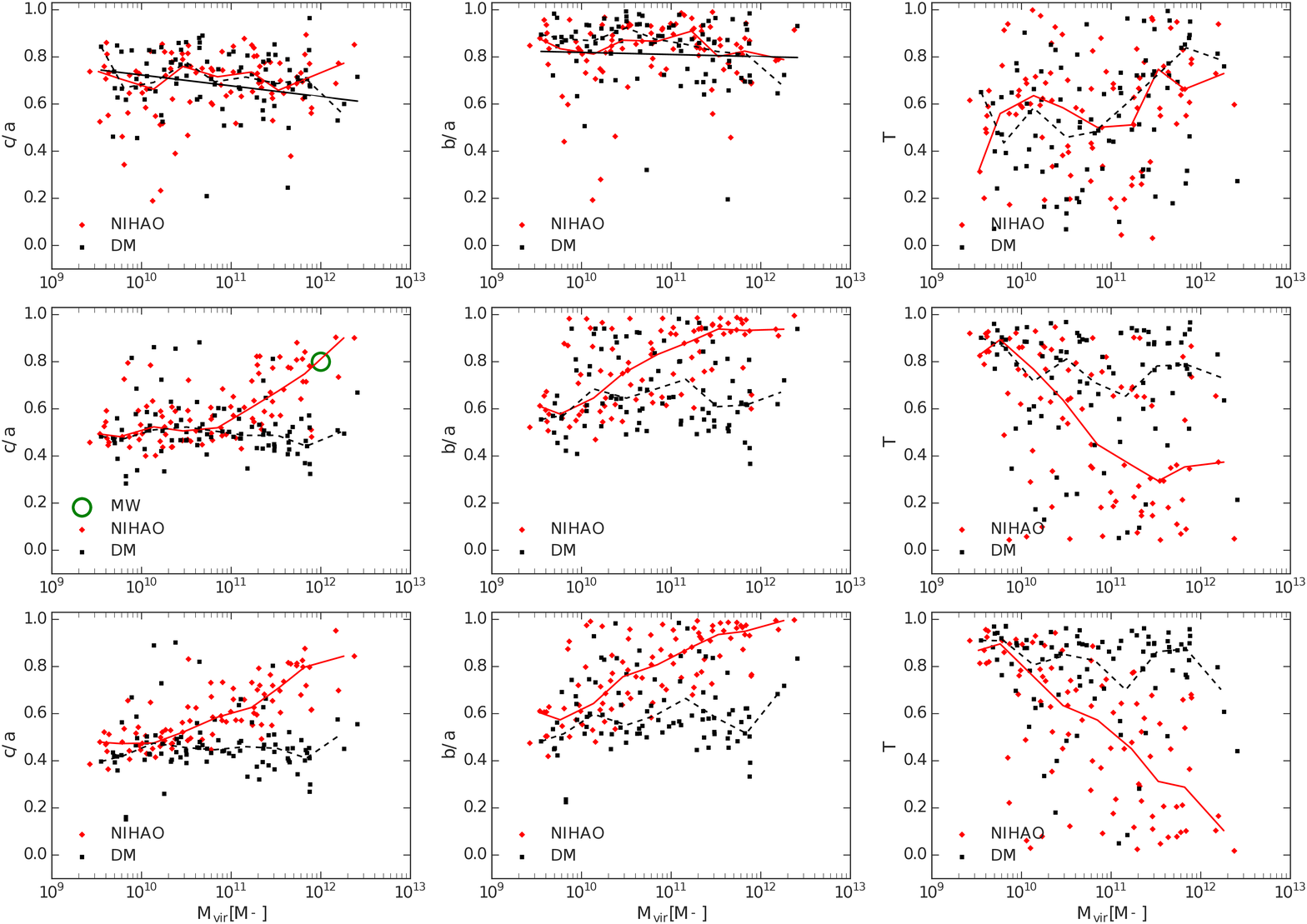,width=1.0\textwidth}}
\caption{\scriptsize Axis ratios ($c/a$, $b/a$) and triaxiality
  parameter ($T$) for all the galaxies as a function of their virial
  (total) mass. The first row shows results at the virial radius, the
  middle row results at 0.12 of \Rvir\, and the last row at 5\% of
  \Rvir. Red diamonds represent the NIHAO galaxies,  while black
  squares show the  corresponding results for the DM-only runs.  The
  solid lines in the first and second panel show the relation from
  \citet{Maccio2008}.  The green circle in the middle left panel shows the
  halo shape measurement for the Milky Way from \citet{Ibata2001}. The red (black)
line shows the median value for the hydro (DM) results.}
\label{fig:shape_mvir}
\end{figure*}

\begin{figure*}
\centerline{\psfig{figure=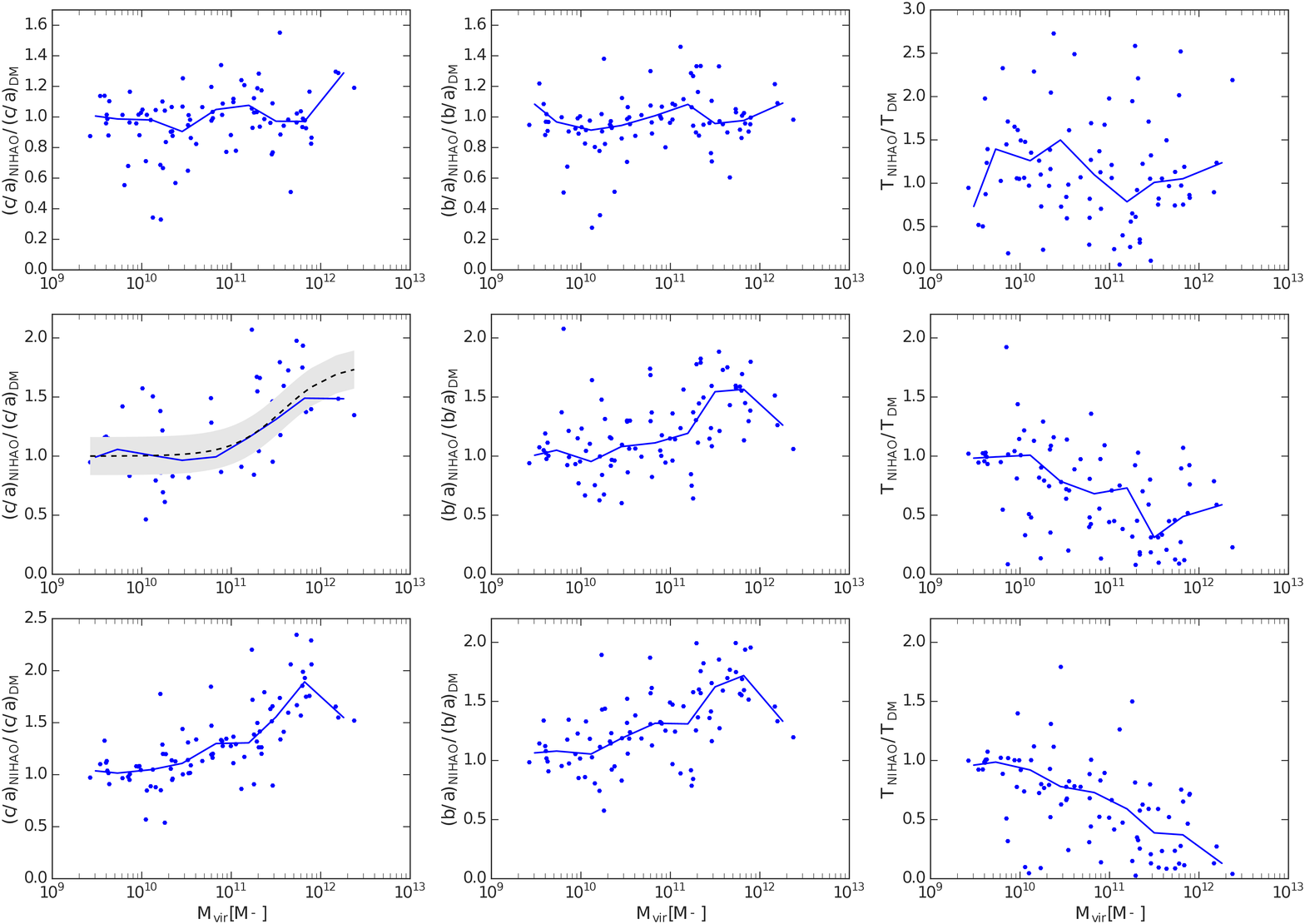,width=1.0\textwidth}}
\caption{\scriptsize Ratio between the  halo shape ($c/a$) in the
  NIHAO and DM-only simulations as a function of the halo mass.  The first
  row shows results at the virial radius, the middle row results at
  0.12 of \Rvir and the last row at 5\% of \Rvir.  The dashed  line in
  the middle-left panel is the fitting function provided in
  Eq.~\ref{eq:sfunc}, the grey area is the $1\sigma$ scatter around
  the mean.  The blue line shows the median value.}
\label{fig:shape_ratio}
\end{figure*}

Fig.~\ref{fig:shape1_all} shows the same quantities as in
Fig.~\ref{fig:shape1} but this time averaged over the four mass bins
M0-M3 (see Table~\ref{tab:gal}), with the grey area representing the
1$\sigma$ scatter around the mean.  The radius at which DM-only and
Hydro results depart from each other moves to larger radii as the mass
grows. In the lowest mass bin (M0, lowest row) there an appreciable
difference between DM-only and Hydro only inside few kpc, while in the
most massive bin (M3, uppermost row) the difference is already
substantial at ~100 kpc.  This is true for all three diagnostics.
Overall the general trends are quite in agreement with the ones from
our test galaxies, which appear to provide a good  representation of
the general population, on the other hand this figure also points out
the not so negligible galaxy-to-galaxy variations in the response of
haloes to baryons.

\citet{Debattista2008} described how the condensation of baryons into
galaxy centers affect the dark matter shape.  DM-only simulations
produce haloes with a prolate shape, $\frac{1}{2} a\sim b\sim c$.
Maintaining the prolate shape relies on dark matter staying on box
orbits.   A prime characteristics of box orbits  is that they make
close passages past the halo center.  As baryons cool and collapse
into the center, they deepen the potential well there and scatter
material from box orbits into rounder loop or tube orbits.  Tube
orbits have an oblate shape, $a \sim b> c$.  Figs.~\ref{fig:shape1} \&
\ref{fig:shape1_all} show the gradual transition of inner halo shapes
from prolate towards oblate as the mass increases:  $b/a$ increases to
1 more quickly than $c/a$.

In order to better quantify the difference in shape between DM-only
and NIHAO simulations on a galaxy-by-galaxy base, we have decided to
compare results for all three shape diagnostics ($b/a$, $c/a$ and $T$)
at three specific radii: $\Rvir$, 12\% $\Rvir$ and 5\%
\Rvir\footnote{The difference in \Rvir between DM-only and Hydro is
  always of the order of few percent and hence does bias our
  results.}.

The virial radius has been chosen to have the possibility to make a
direct comparison with previous studies based on DM-only simulations
\citep{JingSuto2002, Allgood2006, Bett2007, Maccio2008}.  On the other
hand, observationally measuring the shape of dark matter haloes at the
virial radius is a quite difficult task since there are very few (if
any) tracers of the DM shape (or potential) that extend so far from
the galaxy center. We have then decided to look at the halo shape at
$0.12 \times R_{\rm vir}$, which is in the middle of the  20-60 kpc
radius range of the halo that \citet{Ibata2001} state shaped  the
Sagittarius stream.  Finally we also look at the shape in the very
inner part of the halo (5\% \Rvir), which is the most prone to be
affected by baryonic effects.

Fig.~\ref{fig:shape_mvir} shows the value of the three shape
diagnostics as a function of the total mass of the halo, at our three
reference radii.  At the virial radius, at all masses the NIHAO
galaxies and their DM-only counterparts show very similar values for
the axis ratio and Triaxiality, and in agreement with results obtained
from larger samples of simulated dark matter haloes \citep[][black
  solid line in the first and second panel]{Maccio2008}.

At smaller radii (second and third rows) there is a steady increase of
$c/a$ (and $b/a$) with M$_{\rm vir}$ in the NIHAO simulations, which
brings the simulated values in good agreement with the MW observations
of \citet[][green circle]{Ibata2001}.  The last column in
Fig.~\ref{fig:shape_mvir} shows the triaxiality parameter vs halo
mass, confirming that  CDM haloes are typically prolate (black
squares). Interestingly, when galaxy formation is included the full
range of halo triaxialities is possible (red diamonds).

The difference between Hydro and DM-only simulations can be better
appreciated in Fig.~\ref{fig:shape_ratio}, where we show the ratio
between the halo shapes $c/a$ (left), $b/a$ (center) and $T$ (right)
between collisional and collisionless simulations as a function of
halo mass.  The figure shows that there is a consistent shift for the
inner halo shape from the DM-only to the baryon simulation and the
trend of halo shapes becomes more spherical with increasing halo mass
is clearly visible.

For the left middle panel (0.12 \Rvir, the most accessible to
observations), we decided to fit such a ratio with a simple $S$-shape
function:
\begin{equation}
S(M) = s_1 + {{(s_2-s_1)} \over {1+(M/M_0)^\beta}}
\label{eq:sfunc}
\end{equation}
where $M$ is the virial mass of the halo. We fixed the value of the
$s_2$ parameter to 1.0 and fit for the other ones using the
Levenberg-Marquardt method, results are reported in
Table~\ref{tab:sshape}.  The final fitting function is shown by the
black line in Fig.~\ref{fig:shape_ratio}, while the grey area
represents the $1\sigma=0.156$ scatter around the mean.

\begin{table}
\begin{center}
  \caption{Fitting parameters describing the ratio between the halo
    shape, $c/a$, at 0.12 $\Rvir$ in the hydro and DM-only simulations
    (middle left panel of Fig.~\ref{fig:shape_ratio}). }
\label{tab:sshape}
\begin{tabular}{cccc}
\hline
$s_1$ & $s_2$ & $M_0$ [$\Msun$]  & $ \beta$ \\
\hline
1.848 & 1.0 &  $3.1 \times 10^{11}$ & 1.49 \\ 
\hline
\end{tabular}
\end{center}
\end{table}

\begin{figure*}
\centerline{\psfig{figure=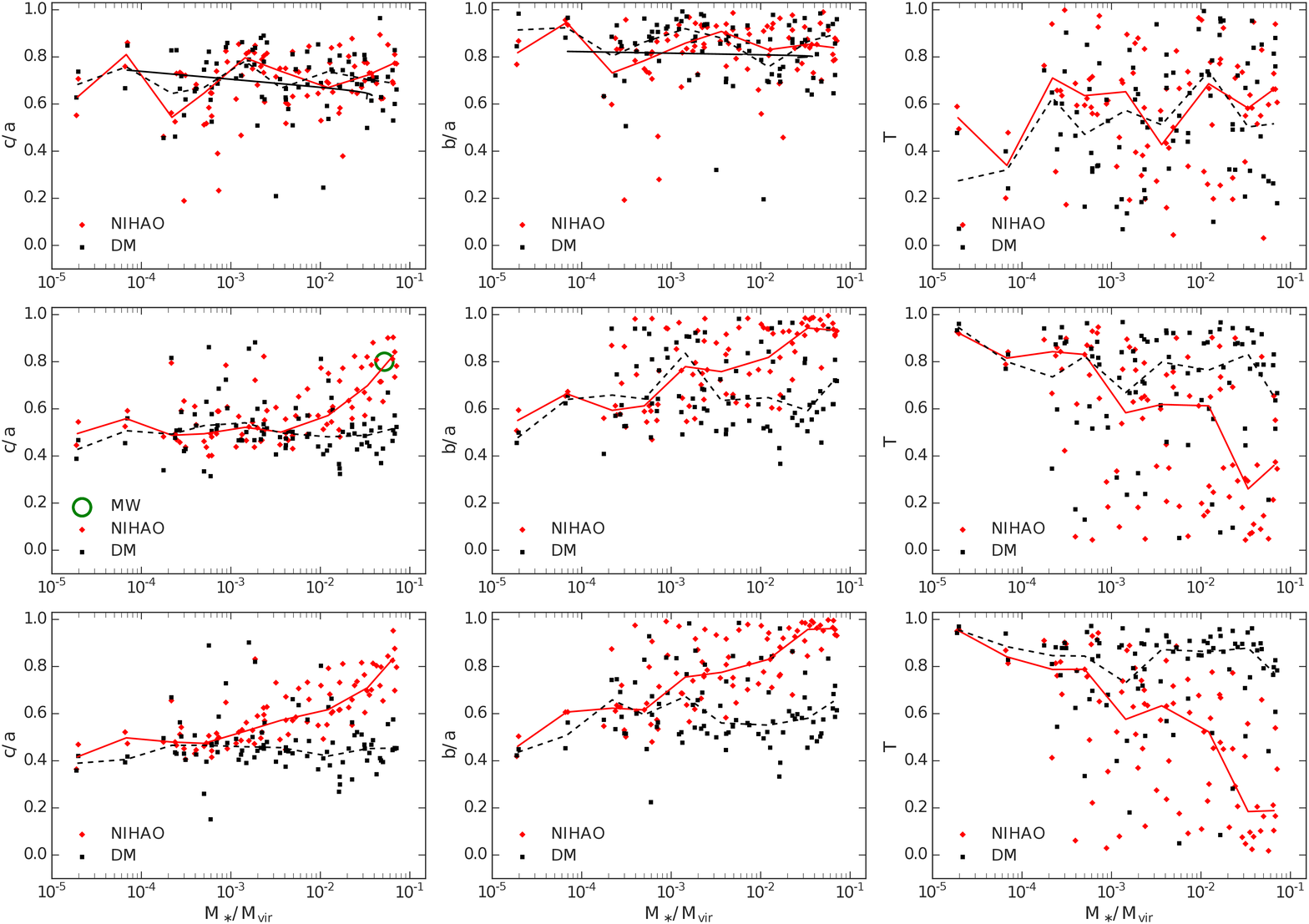,width=1.0\textwidth}}
\caption{\scriptsize Same as Fig.~\ref{fig:shape_mvir} but this time
  as a function of the star formation efficiency  of the halo,
  parameterized as $\Mstar/\Mvir$.  The values for $\Mstar/\Mvir$ for
  the DM-only simulation are obtained using the relation from
  \citet{Moster2010}.  The red (black) line shows the median value for
  the hydro (DM) results.}
\label{fig:shape_ms}
\end{figure*}

The trend of shape change with halo mass can be understood as a
consequence of the increased efficiency of star formation of massive
haloes (in our sample, e.g. $10^{12}$ \Msun) with respect to low mass
ones, as implied by abundance matching \citep{Moster2010}.
Fig.~\ref{fig:shape_ms} shows the different shape measurements versus
the ratio between the stellar and the total mass.  Since there are no
stars in the DM-only simulation we use the empirical formula of
\citet{Moster2010} to assign a stellar mass to each halo (black
squares). 

As expected for a very low values of $M_*/M_{\rm vir}<0.01$ the
influence of baryons is minimal and the halo shape does not change
substantially between DM-only and hydro simulations.  For larger values of
the stellar-to-total mass ratio, the two distributions tend to
diverge, with the NIHAO galaxies showing a substantially rounder halo
shape.

A similar trend  also applies to the Triaxiality parameter (last
columns).  When galaxy formation is very inefficient $(M_{\rm
  star}/M_{\rm vir} < 0.01)$ halo retain their prolate shape.  Above
$(M_{\rm star}/M_{\rm vir} > 0.01)$ haloes can become both triaxial
$T\sim 0.6$ and close to oblate $T\sim 0.2$.  This is quite important
since, when embedded within a triaxial dark matter halo there can be
systematic differences between the rotation speed and the circular
velocity in gaseous discs,  which has important implications for
interpreting observations of dark matter density profiles
\citep{Hayashi2006}.
  
The shape of the halo leaves an imprint on the 2D kinematics
\citep{Kuzio2011}, with triaxial haloes yielding twists along the minor
axis, and spherical (or aligned axisymmetric) haloes yielding
symmetric velocity fields.  Our simulations thus predict a wider
diversity of kinematic structures than would be inferred from DM-only
simulations.

Recently \citet{Kazantzidis2010} have used controlled (N-body)
experiments to study the effect of the growth of a disc onto a
triaxial halo.  They found that the net effect depends weakly on the
time scale of the disc assembly but strongly on the overall
gravitational importance of the disc.

\begin{figure}
\centerline{\psfig{figure=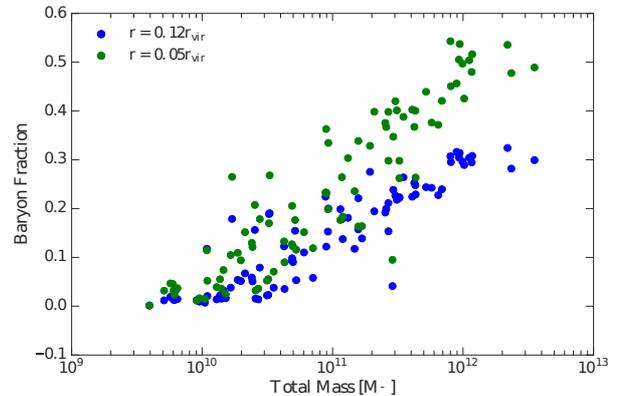,width=0.5\textwidth}}
\caption{\scriptsize Baryon fraction at 12\% (green)  and 5\% (blue)
  of $\Rvir$ as a function of total mass.  It is clear that baryons
  start to be significant only above a total mass of few $10^{11}\Msun$.}
\label{fig:baryon_fraction}
\end{figure}

In order to test their results on the importance of the baryonic mass
(stellar and gaseous)\footnote{Since we have fully cosmological
  simulations we have decided to use the global baryonic contribution
  and not just the disc one.}
contribution to the local  potential in Fig.~\ref{fig:baryon_fraction}
we show the baryonic fraction, $[M_{\rm gas}(R) +M_*(R)]/M_{\rm
  tot}(R)$, computed at R=5\%$\Rvir$ (green) and 12\%$\Rvir$ (blue),
as a function of halo mass.  Let us underline that our galaxies are
very realistic stellar masses (see Fig.~1) and hence our baryonic
fractions should be minimally effected by a possible over-cooling.  A
comparison of this figure with Fig.~\ref{fig:shape_ratio} shows that a
difference between DM-only and hydrodynamical simulations becomes
clear only when the baryonic (mostly stellar) mass is at least of the
order of 10\% of the total one, supporting the findings of
\citet{Kazantzidis2010}.

\begin{figure}
\centerline{\psfig{figure=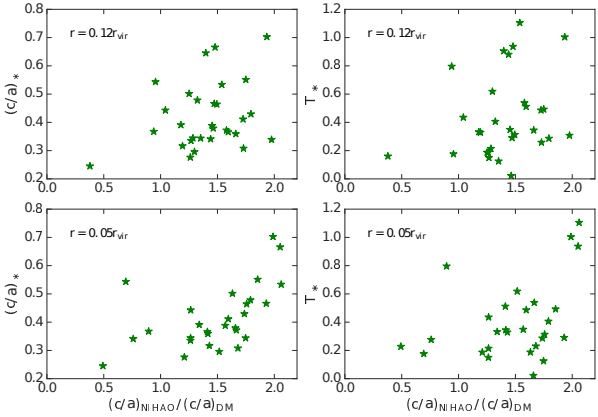,width=0.45\textwidth}}
\caption{\scriptsize Relation between the shape of the stellar
  component and the relative change in the inner shape (12\% $\Rvir$
  upper panel and 5\% $\Rvir$ lower panel) between NIHAO and the
  DM-only run. }
\label{fig:stellar_shape}
\end{figure}

Finally before concluding our section on the DM halo shape we want to
look at the effects of the shape of the stellar component (i.e. the
galaxy morphology) on the final halo shape.
Fig.~\ref{fig:stellar_shape} shows the relation between the shape of
the stellar component (computed at  5\% $\Rvir$, radius that includes
more than 90\% of all stars) and the relative change in the inner
shape (12\% $\Rvir$ upper panel and 5\% $\Rvir$ lower panel) between
NIHAO and the DM-only run. In order to avoid to mix the effect of halo
mass with the effect of stellar morphology this plot has been done
using only galaxies in the M2 mass bin.

On the larger spatial scale the correlation (if present) is very weak,
while a positive trend is more apparent on the very small scale of 5\%
\Rvir.  On this scale it looks that a more spherical stellar
distribution causes stronger change in the DM halo shape. This
correlation can be explained by the fact that rounder stellar bodies
(i.e. bulges) are more compact than disc ones and hence more effective
in modifying the orbit of DM particles \citep[e.g.][]{Debattista2008}.

\subsection{Dark matter pseudo phase-space density}

Moving beyond the morphology of the dark matter, it is possible to
also consider its kinematics.  \citet{Taylor2001} defined ``pseudo phase
space density'' as a simple relationship between  matter density and
velocity dispersion, a quantity that describes the matter distribution
and kinematics of the dark matter together.  They defined  pseudo
phase space density as: 
\begin{center}
$Q(r) = \rho(r) / \sigma^3(r)$,
\end{center}
where $\sigma(r), \rho(r)$ are the velocity dispersion and density of
the halo, respectively.  \citet{Taylor2001} found that  this simple
combination of properties serves as a useful probe for understanding
the origin of the universal DM halo profiles.

Using DM-only simulations \citep{Taylor2001} found that the pseudo
phase space density follows a simple power law, $Q(r) \propto
r^{\chi}$, with $\chi \sim -1.875$.  Since there is mounting evidence
that baryons modify the DM density profile, $\rho(r)$,
\citep[e.g.,][]{Mashchenko2006, Governato2010,
  Teyssier2012,DiCintio2014a,Tollet2016},  it is worth checking
whether baryons also reshape the $Q$ profile.

\begin{figure*}
\centerline{\psfig{figure=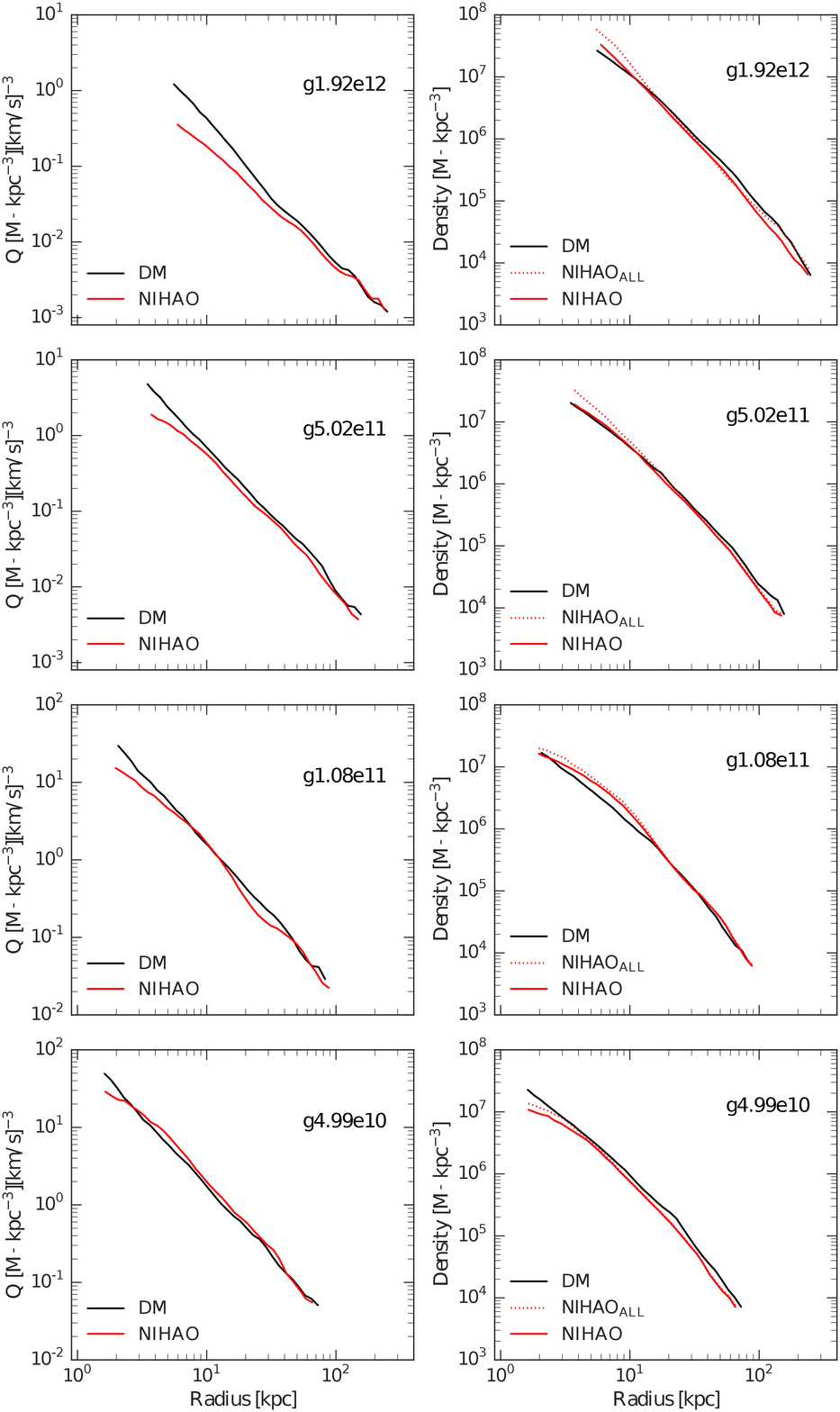,width=0.74\textwidth}}
\caption{\scriptsize The density (left) and pseudo phase space density
  (right)  of DM-only (black) and NIHAO (red) galaxies for our four test
  objects.  Solid lines represent the  density and $Q$ values of the
  dark matter particles, while the red dotted line represents the
  total density of the star, gas, and dark matter particles.}
\label{fig:q}
\end{figure*}

Fig. \ref{fig:q} presents a comparison between the matter density
profile (right) and the pseudo phase-space density $Q$ profile (left)
for our ``test'' galaxies (the same galaxies shown in
Fig.~\ref{fig:shape1}).  Each density profile includes three lines:
the density in the DM-only simulations (black solid line), the DM
density in the hydro simulations (red solid line) and the total
density profile (dark+stars+gas) in the hydro simulations (red dotted
line).  While profiles from DM-only simulations have universal
Einasto-like shapes \citep[e.g.][]{Navarro2004, Merritt2005,
  Dutton2014}, profiles from hydro simulations exhibit a core for low
mass haloes which gradually steepens with halo mass, becoming even
more cuspy DM-only simulations of the most massive galaxies in our
sample (see Tollet \etal 2016 for a thorough discussion of the
modification of the DM density profiles in the NIHAO simulations).

The $Q$ profiles show a different behavior.  $Q$ in hydro simulations
is always lower than its N-body counterpart.  The profiles also
flatten in the center even when the density profiles do not.  The
flattening of $Q$ in the lowest mass halo (bottom panel) reflects the
flattening  of the matter density, which decreases the numerator in
the definition of $Q$. For the other three panels, there is no such
discrepancy in the matter density profile.  However, the ``total''
density profile in the hydro simulation does  get steeper in all three
cases.  The deeper global potential well causes the velocity
dispersion of the DM $\sigma$ to increase towards the center, thus
flattening the inner part of the $Q$ radial profile.  As a consequence
the pseudo phase-space density profile of the Dark Matter component
departs from simple power law behavior at practically all mass scales
probed by our hydrodynamical simulations.

A comparison of Fig. \ref{fig:shape1} and  Fig. \ref{fig:q} also
suggests that the violent gas motions caused by baryons in the center
of low mass galaxies which are believed to be responsible of the
flattening of the central dark matter cusps (e.g.
\citet{Pontzen2012}), also works to  make the dark matter rounder.  In
the low mass g4.99e10, baryons make little change to the outer shape
of the dark matter.  However, the dark matter is almost completely
spherical inside 2 kpc,  the same region in which the density profile
has been flattened.

\begin{figure*}
\centerline{
  \psfig{figure=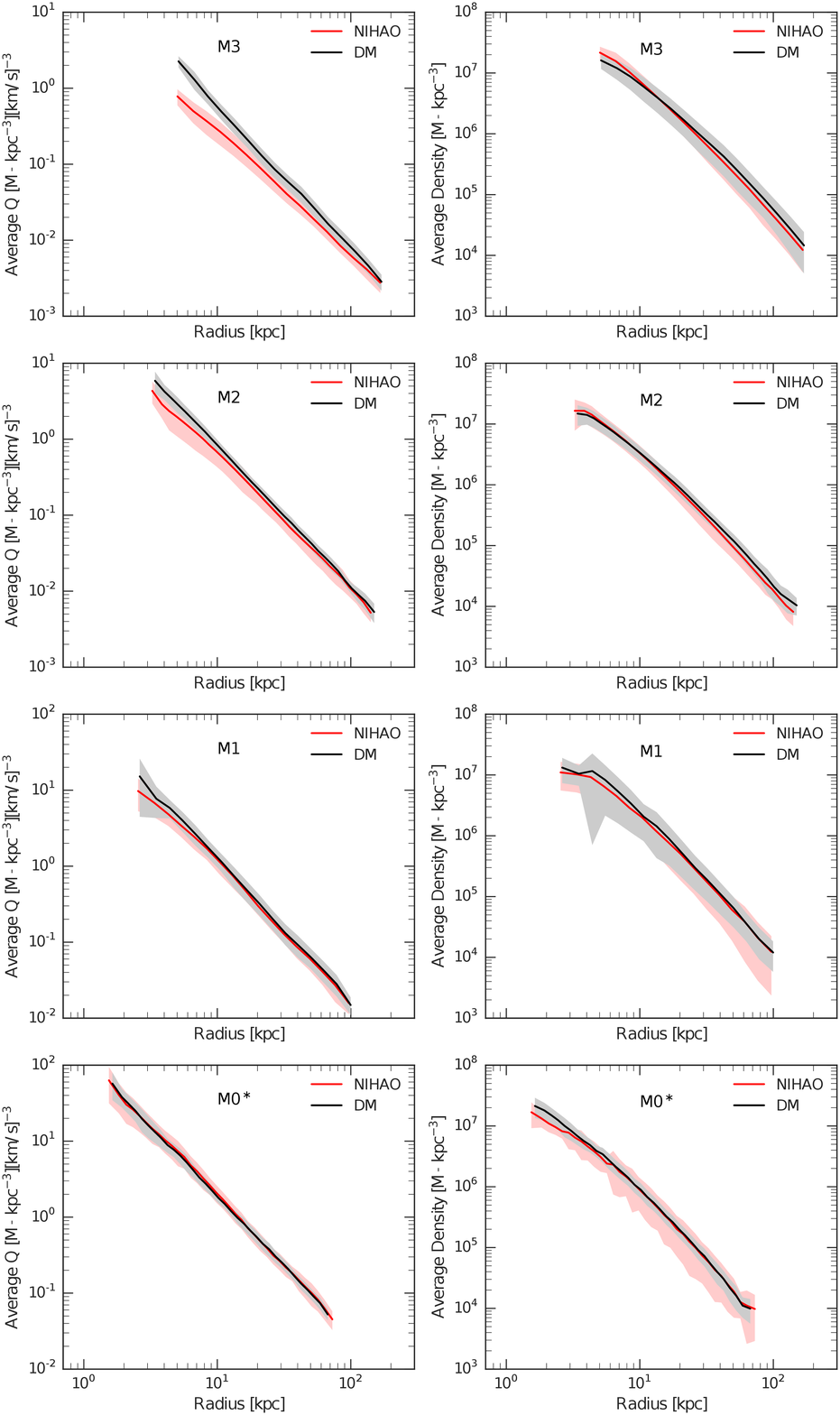,width=0.74\textwidth}}
\caption{\scriptsize The density (left) and pseudo phase space density
  (right)  of DM-only (black) and NIHAO (red) galaxies averaged in the
  for mass bins M0-M3 (M0 $<$ M1 $<$ M2 $<$ M3 $\sim 10^{12}\Msun$).  The
  grey area represents the 1$\sigma$ scatter around the mean.}
\label{fig:qave}
\end{figure*}

Fig.~\ref{fig:qave} shows the same quantities as in Fig.~\ref{fig:q}
but this time averaged for all galaxies in the different mass bins
M0-M3.  The averaged profiles confirm the trends already seen for the
four ``test'' galaxies.  In the lowest mass bin (M0) there is quite
small difference between the averaged $Q$ computed in the DM-only or NIHAO
simulations, suggesting that stochasticity of star formation is quite
strong on those mass scales as already pointed out in works dealing
with the flattening of the density profiles
\citep[e.g.][]{Onorbe2015,Tollet2016}

\begin{figure}
  \centerline{
\psfig{figure=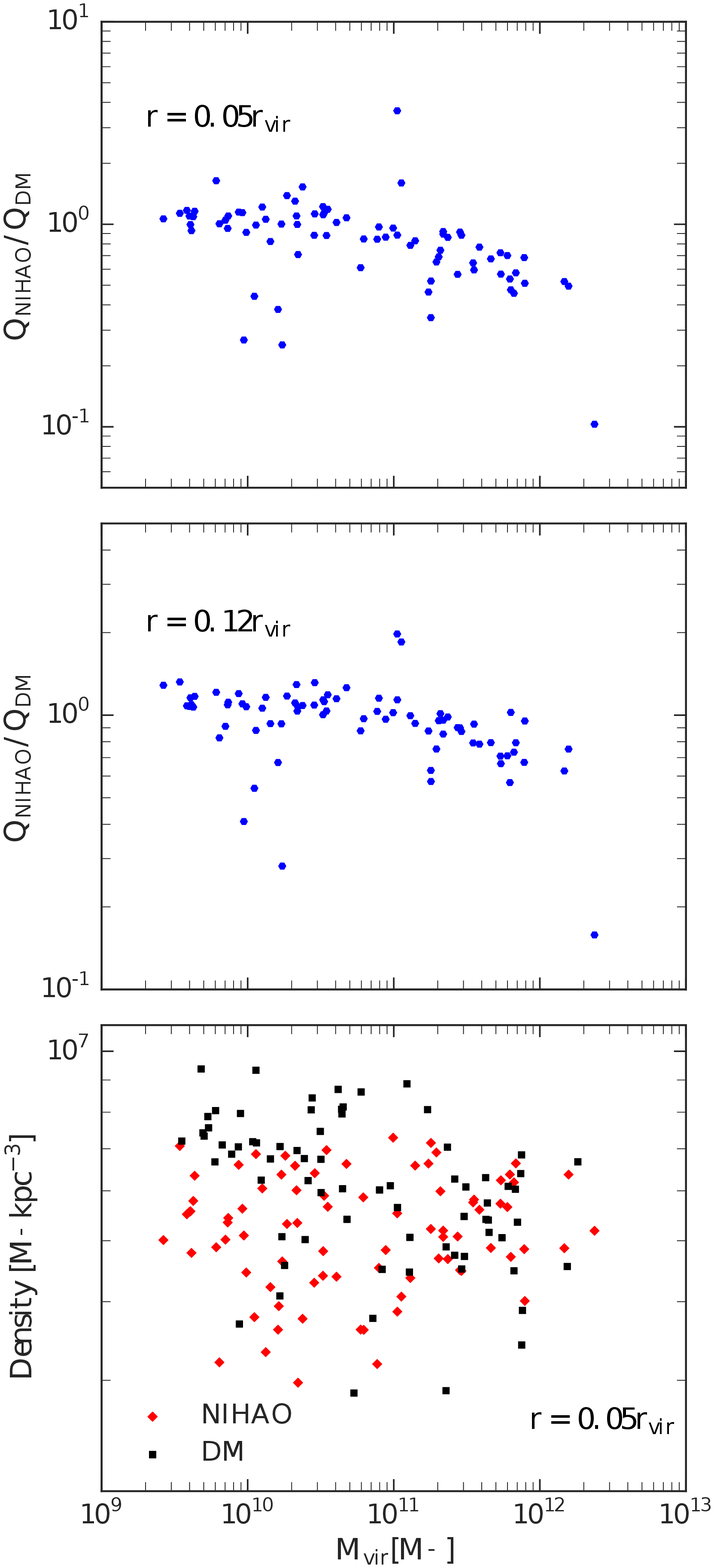,width=0.45\textwidth}}
\caption{\scriptsize The top panel shows the ratio of the values of
  $Q$ in the in dissipationless and hydro simulations computed at 0.05
  \Rvir (0.12 \Rvir for the middle panel).  The outliers in the plot
  are due to merging galaxies (haloes) which are at a  sligtly
  different merging state in the Hydro and DM simulations.  The bottom
  panel shows the values of $\rho$ at 5\% of \Rvir for the DM (black)
  and the NIHAO (red) simulations. }
\label{fig:qall}
\end{figure}

In Fig.~\ref{fig:qall} we try to summarize our findings by showing the
ratio of the  values of $Q$ at two different radii in NIHAO and in
DM-only simulations: 0.05 \Rvir and 0.12 \Rvir. At low masses
(M$\approx 10^{10}$ \Msun) there is no much difference between the two
Qs, at higher masses the hydro simulations show on average a lower
value of $Q$, in agreement with findings based on single profiles.
Most likely this this strong dependence of the $Q$ ratio on halo mass
is due to the $\sigma^3$ term in the denominator of the definition of
the pseudo phase-space density.  The density when computed at 5\% or
\Rvir (lower panel of figure ~\ref{fig:qall}) shows an average lower
value in the hydro simulations up to a mass of few $10^{11}$ solar
masses, and similar values above this mass, consistent with results on
halo expansion and contraction reported in \citet{Tollet2016}.

\subsection{Dark matter velocity distribution}

As outlined in the introduction, the velocity distribution is crucial
for the detection of DM in the Milky-Way. For this reason together
with the results  of our test galaxies, and the corresponding averaged
mass bins (M0-M3), we will also show results for four more single
galaxies namely: g8.26e11,  g1.12e12, g1.77e12, g2.79e12.

These galaxies have been selected in order to have a stellar and halo
mass similar to the one of our own Galaxy ($\approx 10^{12} \Msun$),
three of them (g8.26e11, 1.77e12 and g2.79e12) are strongly disc
dominated with a disc to total ration of $D/T>0.6$ (computed according
to \cite{Obreja2016}), while  g1.12e12 is quite compact with the disc
accounting for $\sim10\%$ of the total stellar mass.  Edge-on and
face-on images of these four galaxies are shown in figure
\ref{fig:4MW}. Each image is 50 kpc on a side and was created using
the Monte Carlo radiative transfer code {\sc sunrise}
\citep{Jonsson2006}.

\begin{figure*}
\centerline{\psfig{figure=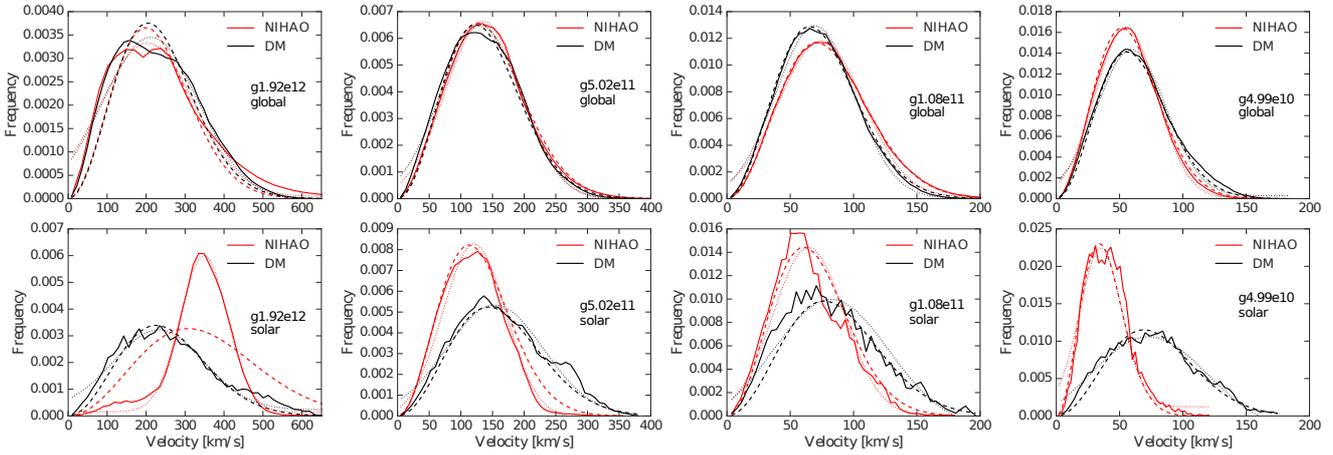,width= 0.99\textwidth}}
\caption{\scriptsize The dark matter particles velocity distribution
  in our four test galaxies.  The dashed lines indicate a Maxwellian
  fit to the simulated curves.  The top panels show the global
  velocity distribution, while the bottom panels show local
  measurements taken at the solar position ( 7 kpc $<$ r $<$ 9 kpc)
  for galaxies in the M3 mass bin, and a similar fraction ($\approx
  5$\%) of the virial radius for lower mass galaxies.  For the local
  velocity  distribution the dotted line shows a Gaussian fit to the
  NIHAO curve.}
\label{fig:vel}
\end{figure*}

Fig. \ref{fig:vel} shows the velocity distribution of dark matter on
two different scales for our test galaxies.  The first row of
Fig. \ref{fig:vel} shows the distribution of all DM particles within
the virial radius while the second row shows the distribution only
inside the solar neighborhood.  This quantity is defined as a shell of
radius between  7 kpc $<$ r $<$ 9 kpc for galaxies in the M3 mass bin,
and it is then rescaled to a similar fraction of the virial radius
(around 4\%) for lower mass galaxies.  In principle the solar
neighborhood should be defined as a ring in the plane of the disc with
radius between 7 and 9 kpc.  We tested on our more disc dominated
galaxies (g8.26e11, g1.77e12, g1.92e12, g2.79e12) that the results do
not change at all if we use a spherical shell instead, which has two
advantages: it increases the number of DM particles (and thus reduces
the numerical noise) and can be applied to both hydro and DM-only
simulations. The similarity of results for a shell and a ring suggests
that the DM distribution is spherically symmetric and no ``dark'' disc
is present in our simulations.  The black lines in figure
\ref{fig:vel} show results from DM-only simulations while red lines
are for the NIHAO galaxies.  At the virial radius, in both types of
simulations, a Maxwellian distribution well represents the global
velocity distribution:
\begin{equation}
f(x) = m_1 \frac{x^2e^{-x^2/(2m_2^2)}}{m_2^3}.
\label{eq:max}
\end{equation}
The best fit Maxwell functions are shown as dashed lines in Fig.~\ref{fig:vel}.

When restricted to the solar neighborhood, the DM-only and the hydro
velocity distributions differ substantially in all  galaxies.  The
DM-only simulations can still be well fit with a Maxwellian
distribution, (even if they show a slightly larger tail at high
velocity).  \citet{Kuhlen2010} found similar agreement  using higher
resolution simulations.  In the hydro case, the velocity distribution
for the most massive test galaxy (g1.92e12) is much more symmetric
around the maximum value, which increases substantially,  and then the
distribution falls quite rapidly at high velocities. This difference
is most likely due to the quite significant halo contraction for this
galaxy (see Fig. \ref{fig:q}) which boosts the local DM velocity
dispersion.

This strong variation at the very high mass end of the NIHAO galaxies
is also confirmed by a closed inspection of the velocity distribution
of the four additional high mass galaxies shown in
Fig.~\ref{fig:vel2}.   With increasing mass the distribution becomes
more Gaussian and the peak moves towards larger velocities, but the
fall-off at the high velocity end is quicker than in the DM-only case.
Despite the different aspect (see figure \ref{fig:4MW}), the four
galaxies have quite similar behavior, suggesting that morphology is
not the main driver of the changes in the velocity distribution, which
seems to be more related to the total stellar (or halo) mass.

For smaller masses (g5.02e11, g1.08e11, g4.99e10) the peak of the
distribution moves towards {\it lower} velocities, possibly related to
the different halo response  on these mass scales, where either
expansion or no reaction (for $M_*/M_{\rm halo}<10^{-4}$) is expected
\citep{Tollet2016}.

\begin{figure*}
  \centerline{\psfig{figure=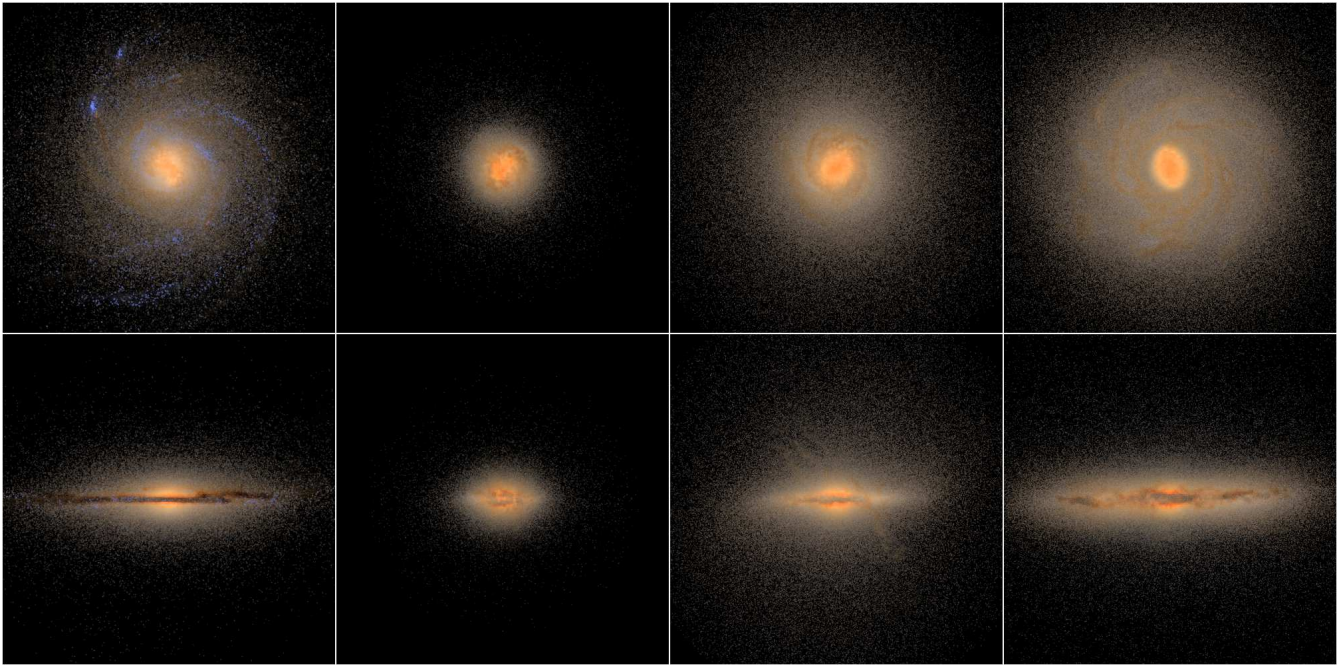,width= 0.99\textwidth}}
\caption{\scriptsize Face-on (upper panels) and edge-on (lower panels)
  views of galaxies: g8.26e11, g1.12e12, g1.77e12, g2.79e12 after
  processing through the Monte Carlo radiative transfer code {\sc
    sunrise}. Images are 50 kpc on a side.}
\label{fig:4MW}
\end{figure*}

\begin{figure*}
\centerline{\psfig{figure=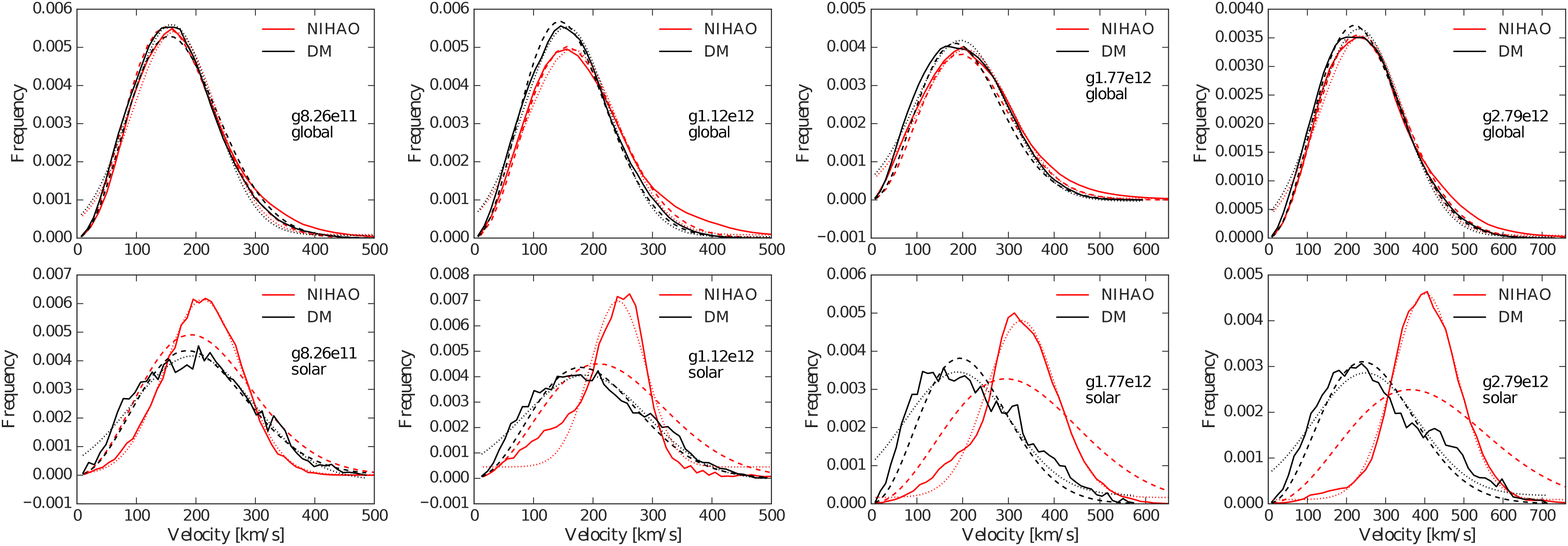,width= 0.99\textwidth}}
\caption{\scriptsize The dark matter particles velocity distribution
  in four additional galaxies with stellar and DM masses similar to
  the Milky Way: g8.26e11, g1.12e12, g1.77e12, g2.79e12.  Symbols and
  lines are the same as in  Fig.~\ref{fig:vel}.  }
\label{fig:vel2}
\end{figure*}

As a consequence of baryonic effects, the local velocity distribution
of MW-like galaxies is better fit by a Gaussian distribution:
\begin{equation}
f(x) =   g_1 e^{-(x-\mu)^2/(2\sigma^2)}
\label{eq:gaus}
\end{equation}
The Gaussian fit is shown in the lower panels of Figs.~\ref{fig:vel}
\& \ref{fig:vel2} by the (red) dotted line that clearly provides a
better fit than a Maxwellian (red) dashed line.  The fitting
parameters for both distributions (Gaussian and Maxwellian) are
reported in Table~\ref{tab:veldist}, all fits have been performed
using the Levenberg and Marquardt algorithm

\begin{table}
\begin{center}
  \caption{Parameters describing the
    velocity distribution function at the virial radius (first two
    rows) and at the solar radius (second two rows) for the five
    Milky-Way like galaxies.}
\label{tab:veldist}   
\begin{tabular}{lccccc}
\hline
\hline
 g8.26e11 & $m_1$ & $m_2$ & $g_1$ & $\mu$ &  $\sigma$ \\
\hline
$R_{\rm vir}$ (dm) & 0.789 & 109.5 & 0.0055 & 158.9 & 70.42 \\
$R_{\rm vir}$ (hydro) & 0.794 & 105.7 & 0.0053  & 164.3  & 70.86\\
$R_{\odot}$ (dm) & 0.776 & 131.0 & 0.0044  & 195.2  & 103.37 \\
$R_{\odot}$ (hydro)& 0.903 & 135.5 & 0.0061 & 211.8 & 64.69\\
\hline
\hline
 g1.12e12 & $m_1$ & $m_2$ & $g_1$ & $\mu$ &  $\sigma$ \\
\hline
$R_{\rm vir}$ (dm) & 0.787  & 101.9  & 0.0055  & 155.6  & 72.28\\
$R_{\rm vir}$ (hydro)  & 0.755  & 110.7 & 0.0049 & 165.0 & 76.96 \\
$R_{\odot}$ (dm) & 0.754 & 127.0 & 0.0041 & 187.3  & 104.2 \\
$R_{\odot}$ (hydro)& 0.905  & 147.7  & 0.0066  & 241.9  & 46.79 \\
\hline
\hline
 g1.77e12 & $m_1$ & $m_2$ & $g_1$ & $\mu$ &  $\sigma$ \\
\hline
$R_{\rm vir}$ (dm) & 0.725  & 129.8  & 0.0042  & 195.5  & 99.00\\
$R_{\rm vir}$ (hydro)  & 0.713  & 137.0 & 0.0039 & 206.6 & 99.80 \\
$R_{\odot}$ (dm) & 0.712 & 137.1 & 0.0032  & 189.2  & 115.22 \\
$R_{\odot}$ (hydro)& 0.929  & 208.3  & 0.0047  & 328.7  & 76.69 \\
\hline
\hline
 g1.92e12 & $m_1$ & $m_2$ & $g_1$ & $\mu$ &  $\sigma$ \\
\hline
$R_{\rm vir}$ (dm) & 0.739 & 144.5 & 0.0035  & 211.6  & 122.3 \\
$R_{\rm vir}$ (hydro)  & 0.699 & 140.37 & 0.0033 & 203.6 & 119.5 \\
$R_{\odot}$ (dm) & 0.735& 159.5 & 0.0030  & 232.0  & 115.2 \\
$R_{\odot}$ (hydro)& 0.969 & 218.0& 0.0058 & 348.1& 60.49\\
\hline
\hline
g2.79e12 & $m_1$ & $m_2$ & $g_1$ & $\mu$ &  $\sigma$ \\
\hline
$R_{\rm vir}$ (dm) & 0.783  & 155.0  & 0.0036  & 231.9  & 110.0 \\
$R_{\rm vir}$ (hydro)  & 0.774  & 160.1 & 0.0035 & 239.4 & 109.2 \\
$R_{\odot}$ (dm) & 0.722 & 170.9 & 0.0027  & 249.0  & 133.8 \\
$R_{\odot}$ (hydro)& 0.870  & 256.6  & 0.0045  & 401.5  & 81.91 \\
\hline
\end{tabular}
\end{center}
\end{table}

\begin{figure*}
\centerline{\psfig{figure=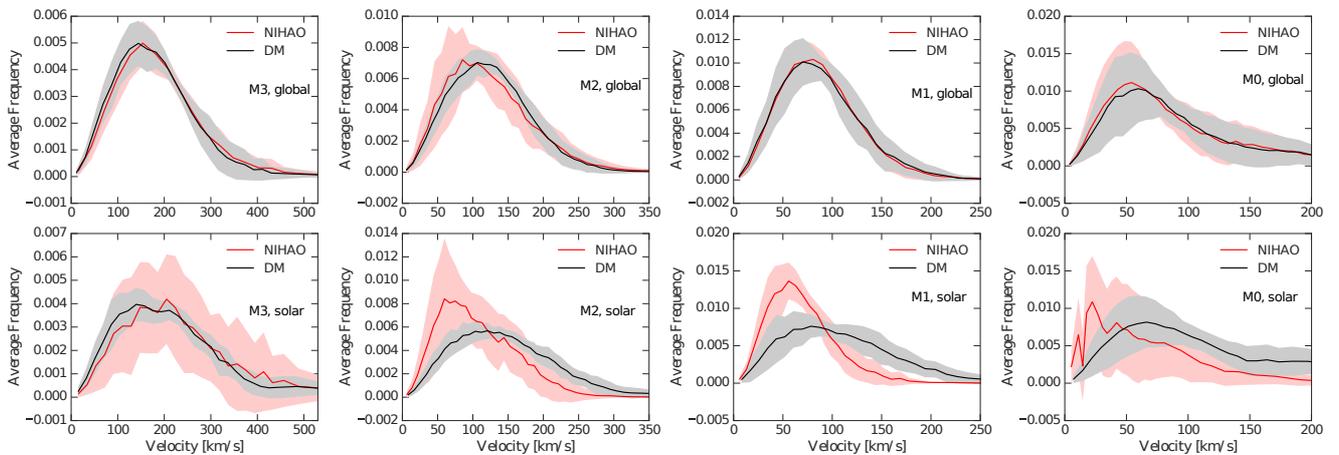,width= 0.99\textwidth}}
\caption{\scriptsize Same as Fig.~\ref{fig:vel} but averaged in the
  four mass bins M0-M3 (M0 $<$ M1 $<$ M2 $<$ M3 $\sim10^{12}\Msun$).}
\label{fig:vel_ave}
\end{figure*}

Finally Fig.~\ref{fig:vel_ave} shows the average velocity distribution
and the virial radius (upper panels) and at the solar radius (lower
panels). While the trend seen is individual galaxies confirmed in the
M0-M2 bins (with g1.99e10 being a bit borderline), this figure also
illustrates the non negligible scatter in the galaxy-to-galaxy
variation for the local velocity distribution.

In the highest mass bin (M3) the results previously highlighted get
washed away due to the large scatter galaxy-by-galaxy as marked by the
extended red area. This scatter is due to the rapid change in halo
response (from expansion to contraction) across this mass bin.  For
lower masses the average values confirm trends seen previously on
single galaxies, namely more symmetric distributions and a lower tail
at high velocities.

On the Milky-Way scale our findings confirm earlier results based on
the single halo ERIS simulation \citep{Guedes2011} and presented by
\citet{Pillepich2014}. In their paper Pillepich and collaborators also
reported a suppression of the wings and a more symmetric shape for the
velocity distribution. In our simulations the differences between the
DM-only case and the hydro simulations seem to be even larger. This
discrepancy might be due to the stronger feedback model (which leads
to a stronger baryonic impact) we have adopted in our simulations in
order to balance the metallicity dependent gas cooling, which was
ignored in the original ERIS simulation.

As extensively discussed in Pillepich et al., the suppression of the
tail of the distribution at high velocities has important consequences
on the interpretation and the comparison of different direct detection
experiments. For example it relaxes the tension between the (possible)
signal of dark matter scattering reported by CDMS-Si \citep{CDMS2013}
and the exclusion of such a signal from the Xenon-100 experiment
\citep{Aprile2012}.  We refer the reader to \citet{Mao2014} and
\citet{Pillepich2014} for a more thorough discussion.

\section{Conclusions}
\label{sec:conclusions}

We used the NIHAO simulation suite \citep{Wang2015} to investigate the
impact of galaxy formation on the properties of the dark matter
distribution within haloes.

The NIHAO suite is a large simulation campaign aiming to produce a
large sample of high resolution simulated galaxies in a cosmological
context. It is an extension of the MaGICC simulations \citep{magic}
and it has been performed with an improved version of the SPH {\sc
gasoline} code (\citealt{Keller2014}) which fixes the well known
problems of particle based hydrodynamical codes \citep{Agertz2007}.
The NIHAO project counts more than 90 simulated galaxies across two
orders of magnitude in halo mass ($10^{10}-10^{12} \Msun$), with each
of the galaxies resolved with at least $4\times 10^5$ elements.

The NIHAO galaxies have been very successful in reproducing stellar to
halo mass ratio on more than five orders of magnitude in stellar mass:
from $10^5$ to $10^{11}\Msun$.  They also show very realistic star
formation histories for their stellar and halo masses.  Finally NIHAO
galaxies are also consistent with the observed gas content of galaxy
discs and  haloes \citep{Stinson2015,Wang2016,Gutcke2016}.  Thanks to
the unprecedented combination of high resolution and large statistical
sample, the NIHAO suite offers a unique set of objects to study the
distribution response of several DM properties to galaxy formation.
Moreover since for each galaxy we have, at the same resolution, an
N-body only  (collisionless) simulation and a full hydrodynamical
simulation, we are able to assess the effect of baryons on a
halo-by-halo basis.

In this study , we focused on three key properties of dark matter
haloes: the halo shape, the radial profile of the pseudo phase-space
density and the dark matter velocity distribution both global and in
the solar neighborhood.  Our results can be summarized as follows:

\begin{enumerate}

\item The shape of the dark matter halo within the virial radius is
  similar between DM-only and hydro simulations. At smaller radii,
  however, the hydro simulations become rounder. There is a strong
  mass dependence to the difference between the inner halo shape
  (measured at 12\% of the virial radius) from the DM-only simulations
  and hydro simulations.  At low masses ($< 10^{11}\Msun $) the dark
  matter halo tends to retain its original triaxial shape, while at
  higher masses ($\approx 10^{12}\Msun $) the inner halo becomes more
  spherical with an average minor to major axis ratio ($c/a$) of 0.8 .
  This brings numerical predictions into good agreement with estimates
  of the inner halo shape in our own Galaxy.  We show that the mass
  dependence of the variation of the halo shape is related to the
  increase of star formation efficiency with halo mass, which raises
  the contribution of stars and gas to the overall potential.

  We provide a simple fitting formula that relates the change in the
  axis ratio $c/a$ with the ratio between stellar mass and halo mass,
  and in principle allows the prediction of the shape of a galaxy DM halo
  from its stellar and  dynamical mass.\\

\item 
  In hydrodynamical simulations the radial behavior of the dark matter
  pseudo phase space-density $Q \equiv \rho/\sigma^3$ is not always
  well represented by a single power law. At  total masses $M\gsim
  10^{11}$ \Msun the $Q$  radial profile shows a flattening towards
  the center of the halo.  This is related to the change in the DM
  density profile which strongly departs from the pure DM results in
  the NIHAO simulations \citep{Tollet2016}.  Overall hydro simulations
  have a lower value of $Q$ compared to the Nbody case, when it is
  measured at a fix fraction of the virial radius. \\

\item The velocity distribution of the dark matter particles {\it
  within the virial radius} in the hydro simulations is still well
  represented by a  Maxwellian distribution, and it is similar to the
  DM-only case at all mass scales.

  When we restrict our analysis to the the solar neighborhood (7 kpc
  $< r<$ 9 kpc) we find that in the hydro simulations the velocity
  distribution functional form  strongly depends on the halo mass.

  At low halo masses ($M\sim 10^{10} \Msun$ the Hydro and DM-only
  simulations show a similar behavior,  when we move to higher masses
  ($M\sim 10^{11} \Msun$) the velocity   distribution becomes
  progressively more symmetric and the velocty peak moves towards
  lower values w.r.t the Nbody case, then in our most massive bin
  ($M\sim 10^{11} \Msun$) the distribution is again in agreement with
  the Nbody case. We tentatively ascribed this trend to the different
  reaction of the DM distribution as function of increasing halo mass
  from few $10^9$ \Msun to M$>10^{12}$ \Msun (namely no effect, halo
  expansion, no effect and  halo contraction) as described in details
  in \cite{Tollet2016}   and \cite{Dutton2016b}.

  To better study this effect, we isolated five galaxies that for
  stellar and total mass resemble our own Milky Way. For these
  galaxies the maximum of  the velocity distribution in the hydro
  simulations moves to {\it higher} velocities w.r.t the Nbody case,
  due to an overall halo contraction in these galaxies, and we find
  very little correlation with the galaxy morphology.  In our Milky
  Way analogs the velocity distribution is well fitted by a  Gaussian
  and we provide the fitting parameters of the distribution for five
  different galaxies.  We also stress that the lack of high  velocity
  particles has important consequences for the interpretation and
  comparison of Dark Matter direct detection experiments.\\

\end{enumerate}

Our results show that baryons have important effects on the dark
matter not only in the very inner part (e.g. leading to expansion and
contraction) but also on the global properties of dark matter.  The
understanding of the nature of dark matter and the comparison of
theoretical predictions with observational data can no longer rely on
pure collisionless simulations, but must include the effects of
visible matter.

\section*{Acknowledgments}

We thanks an anonymous referee whose comments strongly improved the
presentation of our results.
The simulations were performed on the {\sc theo} cluster of the Max
Planck Institute for Astronomy, and {\sc hydra} cluster, both based at
the Rechenzentrum in Garching, and on the High Performance Computing
resources at New York University Abu Dhabi.
AVM, AAD, and GSS  acknowledge support from the
Sonderforschungsbereich SFB 881 ``The Milky Way System'' (subproject
A1) of the German Research Foundation (DFG).
IB contribution to this project was made possible through the SURF
program at Caltech, and was supported by the Flintridge Foundation,
Caltech SFP Office, and Christian Ott. IB also acknowledges support
from the Sonderforschungsbereich SFB 881 ``The Milky Way System''
(subproject A1) of the German Research Foundation (DFG) during her
stay at the MPIA.
XK acknowledge the support from NSFC project No.11333008.
CP is supported by funding made available by ERC-StG/EDECS n. 279954.
Finally IB would like to thank Lynne Hillenbrand for her guidance in
writing this paper.

\bibliographystyle{mnras}
\bibliography{bibliography}

\label{lastpage}

\end{document}